\documentclass[preprint,pra,aps,nofootinbib,longbibliography]{revtex4-1}
\usepackage{graphicx}
\usepackage{braket}
\usepackage{amssymb}
\usepackage{amsmath}
\usepackage{xcolor}
\usepackage[normalem]{ulem}

\begin{document}

\title{Wick rotation derivation for weak values of the density and time-dependent density functional theory}
\author{Russell B. Thompson}
\email{thompson@uwaterloo.ca}
\affiliation{Department of Physics \& Astronomy and Waterloo Institute for Nanotechnology, University of Waterloo, 200 University Avenue West, Waterloo, Ontario, Canada N2L 3G1}
\author{Zarin Tasneem}
\affiliation{Department of Physics \& Astronomy, University of Waterloo, 200 University Avenue West, Waterloo, Ontario, Canada N2L 3G1}
\author{Yves Caudano}
\affiliation{Namur Institute of Structured Matter (NISM) and Namur Institute for Complex Systems (naXys), University of Namur, Rue de Bruxelles 61, B-5000 Namur, Belgium}
\date{\today}

\begin{abstract}
The equations of time-dependent density functional theory are derived, via the expression for a quantum weak value, from ring polymer self-consistent field theory using a mathematical correspondence between time and imaginary time. The imaginary time path integral formalism of Feynman, in which inverse temperature is seen to be a Wick rotation of time, allows one to write the equilibrium partition function of a quantum system in a form mathematically isomorphic with the path integral expression for the dynamics. Therefore the self-consistent field theory equations which are solutions to the equilibrium partition function are Wick rotated back into a set of dynamic equations, which are shown to give an expression for a quantum weak value of the one-particle density. Remarkably, weak values emerge naturally here without being postulated, as an intermediate step before recovering the standard expression for the density. The weak value expression in turn leads to the equations of time-dependent density functional theory. This first-principles derivation does not use the theorems of density functional theory, which are instead applied to guarantee equivalence with standard quantum mechanics. An expression for finite-temperature dynamics is also given, which shows that a ring polymer model for quantum particles holds for time-dependent systems as well as equilibrium situations. Issues arising in time-dependent density functional theory, such as causality, initial state dependence, and $v$-representability, are discussed in the context of the ring polymer derivation.
\end{abstract}

\maketitle

\section{Introduction}
Mathematical correspondences, sometimes called dualities, or, in more restricted cases, isomorphisms, can be important tools in the formulation of physics. They can allow a single system to be viewed in multiple ways, or, following the perspective of Butterfield \cite{Butterfield2021}, provide mappings between different subject matters. Such mathematical mappings have been used in different forms in both foundational physics and applied research both to test hypotheses and to view phenomena from new perspectives.

Feynman used a correspondence between real time and fictitious imaginary time to express a model for superfluidity \cite{Feynman1953b}. Using a path integral approach for quantum statistical mechanics, he wrote the partition function for a single quantum particle of mass $m$ at a temperature $T$ in a potential $U({\bf r})$ as
\begin{equation}
Z(\beta) = \int d{\bf r}_0 \int_{{\bf r}_0}^{{\bf r}_0} \mathcal{D} {\bf r} \exp \left\{ -\int_0^{\beta} \left[\frac{m}{2\hbar^2} \left(\frac{d{\bf r}}{ds}\right)^2 + U({\bf r}(s)) \right] ds    \right\}  \label{part1}
\end{equation}
where $\beta = 1/k_BT$, $k_B$ is Boltzmann's constant and $\hbar$ is Planck's reduced constant. In equation (\ref{part1}), the position of the particle ${\bf r}$ is parametrized in $s$, so that the functional integral over paths denoted by $\mathcal{D} {\bf r}$ explores thermal ``trajectories'' in the  space of position and inverse thermal energy. Feynman obtained this path integral expression by noting the correspondence between inverse temperature and time: the quantum partition function is the Wick rotation of the quantum dynamics, that is, replacing $s=it/\hbar$ with $s=1/k_BT$ in the path integral expression for the dynamics yields the partition function for the particle given by (\ref{part1}). Due to this replacement, (\ref{part1}) is often called an ``imaginary time'' formalism for quantum statistical mechanics \cite{Videla2023}. To complete this mapping, one should notice that, unlike the path integral for the dynamics, imaginary time trajectories must begin and end at the same spatial location ${\bf r}_0$, as denoted by the limits on the path integral in equation (\ref{part1}). For this reason one can refer to the fictitious thermal dimension as ``cyclic imaginary time''. 

From another perspective, equation (\ref{part1}) obeys a practical isomorphism. If $s$ is taken as embedded in 3D space rather than being an independent thermal or time dimension, then equation (\ref{part1}) is identical to the configurational integral of the standard coarse-grained model of a \emph{classical} ring polymer \cite{Matsen2006, Fredrickson2006}. In such a coarse-grained model, the details of the macromolecule are ignored so that polymers are represented as ideal mathematical space-curves ${\bf r}(s)$ in $\mathbb{R}^3$. These space-curves are fractals in that they appear the same at every length scale. For real polymers, this coarse-graining is an approximation that will break down at scales small compared to the contour length. For fundamental particles such as electrons, where no internal structure is known, this model becomes exact within the imaginary-time representation. Thus one can choose to view quantum particles as classical ring contours, but in a thermal-space. This has led to a small industry of quantum simulation techniques, such as path integral Monte Carlo \cite{Ceperley1995}, centroid molecular dynamics \cite{Zeng2014, Roy1999b} and ring polymer molecular dynamics \cite{Althorpe2009, Habershon2013}. In addition to these simulation techniques, the partition function (\ref{part1}) can be solved using polymer self-consistent  field theory (SCFT), which is a mean-field method for the equilibrium classical statistical mechanics of various architectures of polymers \cite{Schmid1998, Matsen2002, Matsen2006, Fredrickson2006, Qiu2006}. Given the isomorphism between the mathematics of ring polymers and quantum particles, the ring polymer SCFT equations can be solved numerically for a variety of equilibrium quantum situations, such as the electron densities of atoms and molecules \cite{LeMaitre2023a, LeMaitre2023b, Sillaste2022}. The ground state quantum particle densities are given by the limit $\beta \rightarrow \infty$, that is, infinitely long polymers, but computationally, a sufficiently large finite value of $\beta$ converges rapidly to ground state conditions to within arbitrary tolerances \cite{LeMaitre2023a, LeMaitre2023b}.

An advantage of using SCFT is that it can be shown to be formally equivalent to Kohn-Sham (KS) density functional theory (DFT). DFT allows one to avoid calculations involving many-body wave functions that have $3N$ degrees of freedom, where $N$ is the number of particles \cite{vonBarth2004}. Instead, the theorems of standard (time-independent) DFT (Hohenberg-Kohn for the ground state \cite{Hohenberg1964}; Mermin for finite temperatures \cite{Mermin1965}) prove that results equivalent to full wave function calculations are obtainable using only the one-particle density $n({\bf r},\beta)$ (for a system at a temperature corresponding to $\beta$)\footnote{\label{note1}The original theorems of Hohenberg and Kohn \cite{Hohenberg1964} link the density to the external potential, but the constrained search methods of Levy \cite{Levy1979} and Lieb \cite{Lieb1983} link the density directly to the wave function -- see for example the introduction sections and general case explanations in references \onlinecite{DelleSite2014, Vignale2001, Vignale2002}.}, making DFT exact in principle. In practice, many-body and quantum exchange effects are included through an exchange-correlation functional of the one-particle density which must be approximated. For SCFT applied to quantum systems, it has been shown that exchange and entanglement correlations can be included exactly\footnote{Exact in principle, but for practicality, approximations would normally be used within SCFT to include exchange effects for many applications.} for both bosons \cite{Thompson2023, Thompson2025} and fermions \cite{Kealey2024} within a classical paradigm. This leaves classical correlations outside the ring polymer picture of quantum mechanics (QM), and these can then be included in an exchange-correlation functional using the same approximations as in standard DFT. The one-particle density of DFT and SCFT exists in $\mathbb{R}^3$ real space instead of the configuration space $\mathbb{R}^{3N}$, making tractable computations that would be impossible otherwise. For this reason, DFT is a central pillar of modern computational chemistry and solid state physics. Since SCFT, and therefore DFT, can be derived from first principles using the partition function (\ref{part1}) rather than through the theorems of DFT as is typically done, these theorems can instead be used in the reverse direction to prove that all predictions of ground state and finite temperature SCFT based on a classical ring polymer model must agree with those of QM \cite{Thompson2022, Thompson2023}.

There are other theorems of DFT, perhaps most importantly the Runge-Gross theorem \cite{Runge1984}. This theorem allows the replacement of the \emph{time-dependent} many-body wave function with the time-dependent density $n({\bf r},t)$. Given the correspondence between quantum particle dynamics and the equilibrium partition function expressed in equation (\ref{part1}), it follows that one must be able to map the equilibrium SCFT equations describing the particle density $n({\bf r},\beta)$ onto the dynamics for $n({\bf r},t)$ through the replacement of the cyclic SCFT equations with non-cyclic ones followed by a Wick rotation $1/k_BT \rightarrow it/\hbar$. That is, replacing the equations for ring polymers (closed, loop trajectories) with those of linear polymers (open, arbitrary trajectories) followed by a Wick rotation. In other words, using Feynman's correspondence in reverse. This is not a new idea; in path integral Monte Carlo, it has been possible to analytically continue the Green's function from imaginary to real time to access real-time dynamics, although results are severely hampered by statistical noise \cite{Fredrickson2023, Dornheim2019}. As SCFT is a mean field theory rather than a simulation, and is free from noise, this prescription can be implemented robustly through SCFT.

In this paper, we perform this operation and find the dynamic SCFT equations for a system of $N$ quantum particles. Our main results are to show that the SCFT-derived dynamic equations involve a quantum weak value for the expression of the density \cite{Dressel2014b} and reduce to the equations of time-dependent density functional theory (TDDFT). Through the Runge-Gross theorem, this is a proof that all predictions of \emph{time-dependent} QM are obtainable through the assumption that equations involving imaginary time are isomorphic with those involving real time.\footnote{\label{note2} For simplicity of presentation, we restrict ourselves in this paper to all predictions of time-dependent QM within the scope of the original Runge-Gross theorem. In principle however, one can go beyond this to include spin, magnetic fields, currents, etc. \cite{Ullrich2025, Vignale2001,Vignale2002, Vignale1987} and time-dependent versions of the Levy-Lieb constrained search approach \cite{Cohen2005}.} In the context of this paper, one can go as far as considering imaginary time to be an element of reality, or alternatively, one can merely consider it to be a fictional calculational tool that is robust in its duality with real time under a Wick rotation. To bolster the element-of-reality perspective, we give evidence that the ring polymer model applies to both temperature-independent and thermal time-dependent cases. In other words, we give the finite temperature expression for the time- and temperature-dependent density $n({\bf r},\beta,t)$ and we show that this can be phrased as a classical ring polymer propagator. In section  \ref{sec-Theory}, we express the density in terms of a quantum weak value and subsequently derive the equations of TDDFT. In section \ref{sec-Discussion} we discuss issues common to derivations of TDDFT including causality, initial state dependence and $v$-representability. We summarize our results and consider possible future work in section \ref{sec-Conclusions}.

\section{Theory}  \label{sec-Theory}
\subsection{Dynamics}
\begin{figure}
\includegraphics[width=0.6\textwidth]{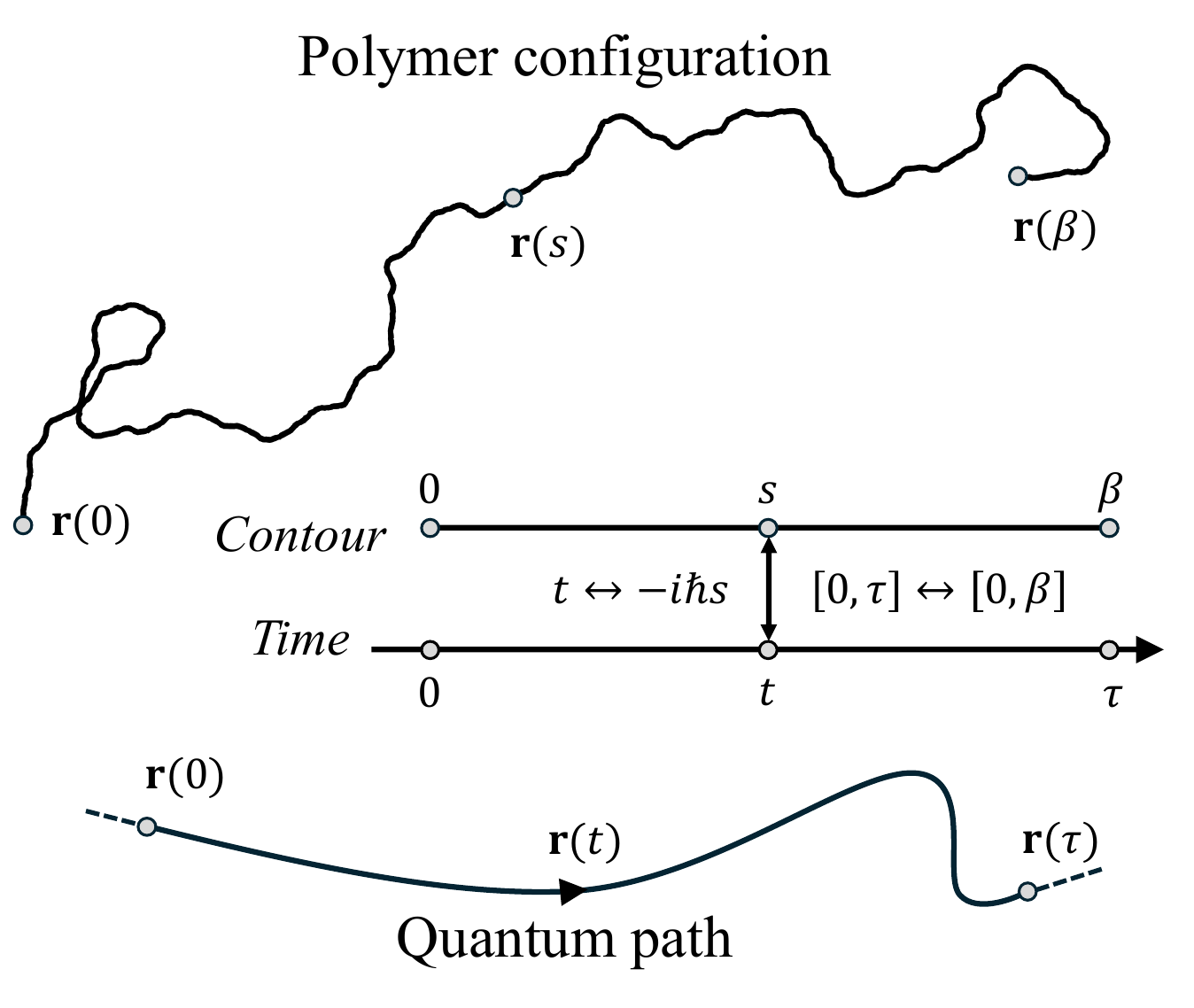}
\caption{Schematic illustrating the difference between (a) a linear polymer represented as an unoriented space curve ${\bf r}(s)$ parametrized from $s=0$ to $s=\beta$ embedded in $\mathbb{R}^3$ and (b) an oriented curve tracing a trajectory ${\bf r}(t)$ \emph{through} $\mathbb{R}^3$ from $t=0$ to $t=\tau$.}
\label{fig-time}
\end{figure}

If the equilibrium statistical mechanics SCFT equations describing a quantum particle in the ring polymer formalism are known, then so are the dynamics, since one is postulating a Wick rotation correspondence between cyclic imaginary time and real time. One need only replace cyclic (ring) polymers in the SCFT equations with linear polymers and likewise replace the $s$ variable for inverse thermal energy with time according to $1/k_B T  \rightarrow it/\hbar$. These changes could be made directly in the partition function to obtain the path integral describing the quantum dynamics, just reversing what Feynman originally did in going from dynamics to statistical mechanics \cite{Feynman1953b}. A standard derivation would then be followed using Hubbard-Stratonovich transformations and the rest of the SCFT mathematics to obtain dynamic expressions \cite{Schmid1998, Matsen2002, Matsen2006, Fredrickson2006, Qiu2006}. It is simpler however just to take the existing SCFT equations and directly transform them to obtain the dynamic equations. Such a process is shown schematically in figure \ref{fig-time}, where the unoriented contour representing a linear polymer is mapped onto an oriented curve parametrized by time.

The SCFT equations for linear polymers are  known and easier to work with than those of ring polymers. Ring polymer SCFT was given relatively recently in papers by Zhang et al. in 2011 \cite{Qiu2011} and Kim et al. in 2012 \cite{Kim2012} whereas linear polymer equations date back to the earliest development of SCFT in the 1960s and 70s (see for example Edwards \cite{Edwards1965} and Helfand \cite{Helfand1975}). Modern and complete expressions for linear polymers can be found in many SCFT review papers \cite{Schmid1998, Matsen2002, Matsen2006, Fredrickson2006, Qiu2006} and, for the convenience of the reader, a very brief summary of both ring and linear polymer SCFT is given in appendix \ref{app-polymers}. For $N$ quantum particles of contour length $\beta$ in equilibrium, the equations for linear polymers can be written as
\begin{eqnarray}
n({\bf r},s) &=& \frac{n_0}{Q(\beta)} q({\bf r},s) q^\dag({\bf r},s) \label{dens1}   \\
Q(\beta) &=& \frac{1}{V}\int q({\bf r},s) q^\dag({\bf r},s) d{\bf r}    \label{Q1}  \\
w({\bf r},s) &=& \frac{\delta U[n]}{\delta n({\bf r},s)}  \label{w1}  \\
\frac{\partial q({\bf r},s)}{\partial s} &=& \frac{\hbar^2}{2m} \nabla^2 q({\bf r},s) - w({\bf r},s) q({\bf r},s)   \label{diff1}
\end{eqnarray}
where $n_0=N/V$ is the average particle density in a volume $V$. The main governing SCFT equation is (\ref{diff1}), which is a modified diffusion equation for a real and non-negative propagator $q({\bf r},s)$ that obeys the initial condition 
\begin{equation}
q({\bf r},0) = 1  .  \label{init1}
\end{equation}
If the polymer is parametrized along its contour from $s=0$ at one end to $s=\beta$ at the other, then the propagator $q({\bf r},s)$ gives the unnormalized probability of finding the infinitesimal piece $s$ of the polymer at the position ${\bf r}$ if starting from the $s=0$ end. The initial condition (\ref{init1}) merely expresses that the $s=0$ end is somewhere in the volume $V$. The total probability of finding $s$ at ${\bf r}$ would be the product of $q({\bf r},s)$ with another propagator $q^\dag({\bf r},s)$ obeying the same equation (\ref{diff1}) but with the left hand side multiplied by $-1$ to encode that it starts from the other side $s=\beta$ and propagates backward along the chain toward $s=0$.\footnote{\label{foot-dag} In SCFT, the dagger symbol $^\dag$ does not mean a mathematical adjoint. It just means that $q^\dag$ is a different real function than $q$. We use the dagger in this work since it is a common notation in polymer SCFT indicating that $q^\dag$ plays a role \emph{analogous} to an adjoint within SCFT.} This second propagator obeys the initial condition
\begin{equation}
q^\dag({\bf r},\beta) = 1  .  \label{init1b}
\end{equation}
Equation (\ref{dens1}) for the inhomogeneous particle density $n({\bf r},s)$ gives the total \emph{normalized} probability of finding the piece $s$ of the chain at position ${\bf r}$. The normalization is given by $n_0/Q$, where $Q(\beta)$ is the partition function of a single polymer subject to the field $w({\bf r},s)$. Note that while the right hand side of (\ref{Q1}) for $Q(\beta)$ appears to be a function of $s$, $Q$ is indeed a function of $\beta$ and not $s$. The reason for this is that for any choice of $s$ on the right hand side of (\ref{Q1}), the product $q({\bf r},s) q^\dag({\bf r},s)$ gives all the possible conformations of the polymer from both sides of $s$ at the position ${\bf r}$. One then integrates over ${\bf r}$ to obtain all possible configurations of the polymer over all possible points in space, and it doesn't matter which point $s$ one chooses to do this with, the final answer is always the same. On the other hand, $Q$ does depend on the length of the polymer $\beta$. More details on the $s$-independence of $Q(\beta)$ can be found in reference \onlinecite{Matsen2002}. $Q(\beta)$ also depends on the field $w({\bf r},s)$ which is determined through equation (\ref{w1}) from the external and internal potentials of the system phrased as functionals of the inhomogeneous density $n({\bf r},s)$. Thus $n({\bf r},s)$ and $w({\bf r},s)$ determine each other self-consistently. For more details, see appendix \ref{app-polymers} or references \onlinecite{Schmid1998, Matsen2002, Matsen2006, Fredrickson2006, Qiu2006}.

A few comments are in order about the differences between equations (\ref{dens1})--(\ref{init1b}) and those of ring polymers. For rings, a propagator $q({\bf r}_0,{\bf r},\beta)$ (and similarly $q^\dag$) is solved to provide the diagonal elements $q({\bf r},{\bf r},\beta)$ needed for the cyclic architecture. But for linear polymers, since the final position need not be the same as the initial position, one can integrate out the ${\bf r}_0$ variable to yield a simpler propagator. Similarly, the initial condition $q({\bf r}_0,{\bf r},0) = \delta({\bf r}-{\bf r}_0)$ for a polymer trajectory can be integrated over ${\bf r}_0$ to yield the simpler initial condition (\ref{init1}) (and likewise (\ref{init1b})). The density for ring polymers as a function of position $n({\bf r})$ has been shown to be (see appendix \ref{app-polymers} and references \onlinecite{Thompson2019, Thompson2020})
\begin{equation}
n({\bf r},\beta) = \frac{N}{Q(\beta)} q({\bf r},{\bf r},\beta) \label{dens1b}
\end{equation}
and so one might expect the linear expression to be
\begin{equation}
n({\bf r},\beta) = \frac{n_0}{Q(\beta)} q({\bf r},\beta)   \label{dens2}
\end{equation}
which is indeed correct. However, unlike in equilibrium SCFT, we shall ultimately be interested not only in the final value of the contour $s=\beta$, but rather in any arbitrary intermediate value of $0 < s < \beta$. Without loss of generality, one can write equation (\ref{dens2}) as equation (\ref{dens1}) for arbitrary $s$ values.\footnote{In polymer physics, one will typically also integrate (\ref{dens1}) over $s$ to get the total monomer density, but this is not a quantity we are interested in for this application and so no $s$ integration is necessary.} One can verify that for the choice $s=\beta$ one recovers (\ref{dens2}) from equation (\ref{dens1}). Similarly, for $s=\beta$, the single particle partition function $Q$ in equation (\ref{Q1}) becomes just
\begin{equation}
Q(\beta) = \frac{1}{V}\int q({\bf r},\beta) d{\bf r}    \label{Q12}  
\end{equation}
as expected. This illustrates again that the partition function value for $Q(\beta)$ does not depend on $s$ even though $s$ appears on the right hand side of equation (\ref{Q1}). Equations (\ref{w1}) and (\ref{diff1}) have also been generalized. For some polymer situations, the field $w({\bf r})$ does not depend on the contour value $s$, but for others it does. For example, for multiblock copolymers, the field in the diffusion equation changes for different chemical species in the chain \cite{Matsen1994b}. For the equilibrium quantum situation, we were only interested in the particle density at one temperature corresponding to $\beta$. For dynamics however, we will be interested in all contour values in order to describe the evolution of the system, and so the more general expressions (\ref{dens1})--(\ref{diff1}) are used. The derivation of these equations is however fundamentally the same.

Now that the SCFT equations have been written for linear polymers instead of ring polymers, we can substitute real time for imaginary time, that is, we replace everywhere $s$ with $it/\hbar$ and likewise $\beta$ with $i\tau/\hbar$, where we will take $\tau$ to be an arbitrary time interval. In other words, we are replacing the linear polymer contour $0 < s < \beta$ with the time interval $0 < t < \tau$. This gives us the equations
\begin{eqnarray}
n({\bf r},it) &=& \frac{n_0}{Q(i\tau)} q({\bf r},it) q^\dag({\bf r},it) \label{dens3}   \\
Q(i\tau) &=& \frac{1}{V}\int q({\bf r},it) q^\dag({\bf r},it) d{\bf r}    \label{Q13}  \\
w({\bf r},it) &=& \frac{\delta U[n]}{\delta n({\bf r},it)}  \label{w3}  \\
\hbar \frac{\partial q({\bf r},it)}{\partial (it)} &=& \frac{\hbar^2}{2m} \nabla^2 q({\bf r},it) - w({\bf r},it) q({\bf r},it)   \label{diff3}
\end{eqnarray}
where the constant $\hbar$ need not be written explicitly in the function arguments. We nonetheless keep $i$ explicit as a reminder that previously real functions are now, in general, complex. As before, $q^\dag({\bf r},it)$ obeys equation (\ref{diff3}) with the left hand side multiplied by $-1$. Possible potentials $U[n]$ in equation (\ref{w3}) will be completely different from those used in polymer physics and must be chosen based on the model of the quantum system being studied. Likewise the initial conditions (\ref{init1}) and (\ref{init1b}) will no longer be relevant and must be replaced by system-specific initial conditions. It is only the \emph{structure} of equations (\ref{dens3})--(\ref{diff3}) that is preserved in the isomorphism; the physical situation is completely different.

\subsection{Weak values} \label{sec-weak}
Equations (\ref{dens3}) and (\ref{Q13}) can be rewritten in a suggestive form. If we define unnormalized functions $\psi_i({\bf r},t) \equiv q({\bf r},it)/\sqrt{V}$, where $t$ is the time parameter, and $\psi_f^*({\bf r},t) \equiv q^\dag({\bf r},it)/\sqrt{V}$, then the density expression (\ref{dens3}) becomes
\begin{equation}
n({\bf r},it) = \frac{N \psi_i({\bf r},t) \psi_f^*({\bf r},t)}{\int d{\bf r}\psi_i({\bf r},t) \psi_f^*({\bf r},t)}  \label{weak1}
\end{equation}
where the exact dependence on $t$ and $\tau$ will be examined later on.
The subscripts $i$ and $f$ denote initial and final states, respectively. Putting equation (\ref{weak1}) in Dirac notation makes explicit that it takes the form of a \emph{weak value}
\begin{equation}
\mathcal{O}_w = \frac{\braket{\psi_f(t)|\hat{\mathcal{O}}|\psi_i(t)}}{\braket{\psi_f(t)|\psi_i(t)}} \label{weak2}
\end{equation}
where the operator $\hat{\mathcal{O}}$ corresponds to $N$ times the projector on the position $\vert \mathbf{r}\rangle$ 
\begin{equation}
\hat{\mathcal{O}} = N \ket{{\bf r}}\bra{{\bf r}}.  \label{weak3}
\end{equation}
$\mathcal{O}_w$ defines a weak value at time $t$ for the particle density, corresponding to the weak value of the position projector multiplied by the particle number \cite{Aharonov1988, Dressel2014b}. Most generally, weak values are unbounded complex numbers. Nevertheless, they yield standard expectation values when $\vert \psi_f(t)\rangle=\vert \psi_i(t)\rangle$. They occur in many quantum contexts, such as representing classical background fields \cite{Dressel2014a}, quantifying interference phenomena \cite{Dressel2015} and as best estimates of observables given known initial and final conditions \cite{Dressel2015}. Weak values also arise in the theory of weak measurements and are used in quantum metrology to obtain large amplification of experimental results or to measure quantum wave functions \cite{Svensson2013, Steinberg2013, Dressel2014b, Lundeen2011}. The theory of quantum measurement is normally essential in applications of weak values as they arise in weak measurements, but there is no theory of measurement given or required in the present derivation since it is based on classical statistical mechanics. In classical physics, measurement can be viewed in an idealized way compared to QM \cite{Greenberger2009}.

The expression (\ref{weak2}) of the density after Wick rotation is best viewed as a conditional quantity that does not emerge from a pre- and post-selection measurement protocol but from a boundary-value problem. In particular, it should be emphasized that the final state $\ket{\psi_f}$ is \emph{not necessarily} the unitary evolution of the initial state $\ket{\psi_i}$. There can potentially be different boundary conditions on $\ket{\psi_i}$ and $\ket{\psi_f}$. The  corresponding argument of the complex weak value carries geometric information related to the boundary conditions and the operator \cite{Ferraz2022, Cormann2017,Ferraz2025}.

In order to understand the significance of the weak value here, it is worth getting back to the polymer picture. The expression for the density (\ref{dens1}) is a conditional probability that results from a rightward diffusion process from the lefthand polymer end $s=0$ (see figure \ref{fig-time}), combined with a leftward diffusion process from the righthand polymer end $s=\beta$. The density contributed at $\mathbf{r}$ by any polymer point $s$ emerges from the product of the probabilities that these two diffusion processes put the polymer segment $s$ at position $\mathbf{r}$. The use of two boundary conditions together with a forward and a backward diffusion process is what allows one to connect the statistical polymer picture to a unitary quantum evolution in the time-dependent case. This mathematical structure is reminiscent of Nelson's stochastic approach to quantum mechanics \cite{Nelson1966, Bacciagaluppi2005}, whereby Schrödinger's unitary dynamics can be recovered by positing forward and backward stochastic drifts associated with a diffusion process under the assumption of time-reversal invariance of the classical dynamical law. However, in the polymer picture, after Wick rotation, the conditioning from the past and future appears naturally, as a reinterpretation of a natural geometric constraint: the propagation of the boundary conditions. The emergence of the weak value (\ref{weak1}) is therefore a direct consequence of formulating the density from a two-boundary (forward–backward) evolution rather than from an initial-value problem. The path followed for the derivation of equations (\ref{dens3})--(\ref{diff3}) yields naturally a formulation of quantum evolutions constrained by initial and final boundary conditions, where weak values play thus a central role. In principle, one could try to work directly with the complex conditional density (\ref{dens3}) appearing in these self-consistent equations and formulate a two-boundary density-functional problem (in which non-trivial boundary constraints may be associated with an effective non-unitary evolution). However, no such generalized density-functional framework is currently available, and developing this approach for practical computation, while clearly of great interest, is beyond the scope of the paper. Instead, we focus on connecting equations (\ref{dens3})--(\ref{diff3}) to standard time-dependent density functional theory.

In the next section, we show how to recover a real density from the weak value. For this, we first need to write precisely the time dependence of the generic weak value (\ref{weak2}). Using Dirac's bra-ket notation, for a polymer with endpoints $\mathbf{r}_0$ and $\mathbf{r}_\beta$, the two propagators providing the density become $q(\mathbf{r}_0,\mathbf{r},s)\equiv \langle \mathbf{r}\vert \hat{T}(s,0)\vert\mathbf{r}_0\rangle$ and $q^\dagger(\mathbf{r_\beta},\mathbf{r},s)\equiv \langle \mathbf{r}_\beta\vert \hat{T}(\beta,s)\vert\mathbf{r}\rangle$, where $\hat{T}(s_2,s_1)$ denotes the evolution operator from the contour values $s_1$ to $s_2$. As expected, the associated partition function is independent of $s$, since $Q(\beta)\propto \int d\mathbf{r}\, \langle \mathbf{r}_\beta\vert \hat{T}(\beta,s)\vert\mathbf{r}\rangle \langle \mathbf{r}\vert \hat{T}(s,0)\vert\mathbf{r}_0\rangle=\langle \mathbf{r}_\beta\vert \hat{T}(\beta,0)\vert\mathbf{r}_0\rangle$, due to the semi-group evolution. After Wick rotation $s\rightarrow i t/\hbar$ and $\beta \rightarrow i \tau/\hbar$, the density becomes $n(\mathbf{r},i t)\propto \langle \mathbf{r}_\tau\vert \hat{T}( i \tau, i t)\vert\mathbf{r}\rangle \langle \mathbf{r}\vert \hat{T}(i t, 0)\vert\mathbf{r}_0\rangle$. Taking into account that the evolution operators are now unitary, i.e. $\hat{T}(i t_2, i t_1)\equiv \hat{U}(t_2,t_1)$, we further simplify the notation to $n(\mathbf{r},i t)\propto \langle \mathbf{r}_\tau\vert \hat{U}(\tau,t)\vert\mathbf{r}\rangle \langle \mathbf{r}\vert \hat{U}(t,0)\vert\mathbf{r}_0\rangle$. Finally, introducing quantum boundary conditions $\psi_f^*(\mathbf{r}_\tau,\tau)=\langle\psi_f(\tau)\vert\mathbf{r}_\tau\rangle$ and $\psi_i(\mathbf{r}_0,0)=\langle\mathbf{r}_0\vert\psi_i(0)\rangle$, we obtain the density expressed as the weak value
\begin{eqnarray}
n(\mathbf{r},i t)&=& N \frac{\iint d\mathbf{r}_\tau\, d\mathbf{r}_0\ \langle\psi_f(\tau)\vert\mathbf{r}_\tau\rangle\langle \mathbf{r}_\tau\vert \hat{U}(\tau,t)\vert\mathbf{r}\rangle \langle \mathbf{r}\vert \hat{U}(t,0)\vert\mathbf{r}_0\rangle\langle\mathbf{r}_0\vert\psi_i(0)\rangle}
{\iiint d\mathbf{r}\,d\mathbf{r}_\tau\, d\mathbf{r}_0\ \langle\psi_f(\tau)\vert\mathbf{r}_\tau\rangle\langle \mathbf{r_\tau}\vert \hat{U}(\tau,t)\vert\mathbf{r}\rangle \langle \mathbf{r}\vert \hat{U}(t,0)\vert\mathbf{r}_0\rangle\langle\mathbf{r}_0\vert\psi_i(0)\rangle}\\
&=&N \frac{\langle\psi_f(\tau)\vert \hat{U}(\tau,t)\vert\mathbf{r}\rangle \langle \mathbf{r}\vert \hat{U}(t,0)\vert\psi_i(0)\rangle}
{\langle\psi_f(\tau)\vert\hat{U}(\tau,t)\hat{U}(t,0)\vert\psi_i(0)\rangle}.\label{WVtdep1}
\end{eqnarray}
The time dependence of the weak value expression (\ref{WVtdep1}) corresponds to the usual expression arising from a weak measurement of an operator $\hat{\mathcal{O}}=N \vert\mathbf{r}\rangle\langle\mathbf{r}\vert$, with pre-selection in state $\vert\psi_i(0)\rangle$ at initial time 0, followed by a weak interaction of the operator $\hat{\mathcal{O}}$ of the system with a meter at time $t$, followed by post-selection in the state $\vert\psi_f(\tau)\rangle$ at time $\tau$. Because the standard weak value expression (\ref{weak2}) appears deceptively independent of $\tau$ at first sight, we note that the corresponding denominator $\langle\psi_f(\tau)\vert\hat{U}(\tau,0)\vert\psi_i(0)\rangle$ of (\ref{WVtdep1}) does not actually depend on the arbitrary intermediate time $t$ but normally depends instead on the interval duration $\tau$, through $\hat{U}(\tau,0)$, considering that the two boundary conditions $\vert\psi_i(0)\rangle$ and $\vert\psi_f(\tau)\rangle$ are arbitrary.\footnote{The dependence of the weak value on the post-selection time $\tau$ must be tracked explicitly when dealing with dissipative processes in open quantum systems \cite{BallesterosFerraz2024} for example, but is otherwise more typically kept implicit.}

\subsection{TDDFT}  \label{sec-TDDFT}

From equation (\ref{WVtdep1}), we can see that only for the final boundary condition $\vert\psi_f(\tau)\rangle=\vert\psi_i(\tau)\rangle$ does the weak-value expression give a real quantity for all times $t$ given an arbitrary initial condition $\vert\psi_i(0)\rangle$ (because $\langle\psi_i(\tau)\vert\hat{U}(\tau,t)=\langle\psi_i(t)\vert$) and $\hat{U}(t,0)\vert\psi_i(0)\rangle=\vert\psi_i(t)\rangle$), and thus this is the only case where it systematically corresponds to a density. For this choice, the final boundary condition corresponds to the initial state evolved to time $\tau$, so that $q^\dagger(\mathbf{r},i t)=q^*(\mathbf{r},i t)$ and, with this specific choice of final boundary condition, the $\tau$-dependence disappears in practice when working with equations (\ref{dens3})--(\ref{diff3}). This last relationship enables computing the density as an initial value problem (like in TDDFT) instead of a two-boundary value problem (as in the linear polymer SCFT). As a result, equations (\ref{dens3}) and (\ref{Q13}) become
\begin{eqnarray}
n({\bf r},it) &=& \frac{n_0}{Q} q({\bf r},it) q^*({\bf r},it) \label{dens4}   \\
Q &=& \frac{1}{V}\int q({\bf r},it) q^*({\bf r},it) d{\bf r}, \label{Q14}  
\end{eqnarray}
where $Q$ no longer depends on the interval $\tau$ (and of course does not depend on $t$ due to unitarity). So we can rewrite equation (\ref{dens4}) as
\begin{equation}
n({\bf r},t) = N |\phi({\bf r},t)|^2 \label{dens5}   
\end{equation}
where we have defined
\begin{equation}
\phi({\bf r},t) \equiv \frac{q({\bf r},it)}{\sqrt{QV}}  .  \label{phi1}
\end{equation}
The form of (\ref{dens5}) requires the quantum particle density $n({\bf r},t)$ to be real, and so we drop the $i$ from its argument. Rearranging equation (\ref{Q14}) as
\begin{displaymath}
\frac{\int q({\bf r},it) q^*({\bf r},it) d{\bf r}}{QV} = 1  
\end{displaymath}
and using the definition (\ref{phi1}) gives the expected normalization condition on $\phi({\bf r},t)$ 
\begin{equation}
\int d{\bf r} |\phi({\bf r},t)|^2 = 1  . \label{phi_norm}
\end{equation}
In DFT, the complex function $\phi({\bf r},t)$ given by (\ref{phi1}) is called an \emph{orbital}, and it is a function of a single position, in contrast to a wave function that is a function of $N$ positions. For fermions, one expects the Pauli exclusion principle to apply, and so all the particles should not coalesce into a single orbital $\phi({\bf r},t)$ as in equation (\ref{dens5}). The origin of the Pauli exclusion principle in the ring polymer formulation of QM will be discussed in section \ref{sec-Discussion}, but operationally, it has been shown that a separate propagator $q_j({\bf r},s)$ can be defined for each particle $j$ (or pair of particles for spin-unpolarized systems assuming spin 1/2), resulting in a separate density following (\ref{dens5}) for each particle (or pair) \cite{Thompson2020}. The total density is then the sum over all particle densities (ignoring spin or for a spin polarized system) and would be
\begin{equation}
n({\bf r},t) = \sum_{j=1}^N |\phi_j({\bf r},t)|^2 \label{dens6}   
\end{equation}
which is the expression for the time-dependent Kohn-Sham density in TDDFT. In this formula, all particles see the same effective potential, just as in TDDFT, such that a single KS potential $w({\bf r},t)$ results from the total density in equation (\ref{w3}) rather than a separate $w_j({\bf r},t)$ for each orbital. The exchange-correlation potential is included in $w({\bf r},t)$. From equation (\ref{dens6})  we see that only the choice of final boundary condition $\vert\psi_f(\tau)\rangle=\vert\psi_i(\tau)\rangle$ gives a real value for the density whereas other choices give complex weak values that are functions of two times\footnote{\label{note7}While in this work we designate the forward time direction preferentially, our formalism is compatible with other approaches such as the two state vector formalism \cite{Aharonov1964, Aharonov1988} where weak values emerge from time propagating backward from the final state and forward from the initial state.}, $t$ and (implicitly) $\tau$, and that depend on an independent final boundary condition. In the configuration recovering a real density, the forward and backward propagators of the two-boundary formulation provide redundant information, so that the density is determined by the evolution of a single set of orbitals from initial conditions, as in standard TDDFT.

TDDFT also obeys the time-dependent KS equation, which is given by equation (\ref{diff3}). Rearranging and writing (\ref{diff3}) for each $\phi_j({\bf r},t)$ while dividing by $\sqrt{Q_jV}$ gives
\begin{equation}
i\hbar \frac{\partial \phi_j({\bf r},t)}{\partial t} = -\frac{\hbar^2}{2m} \nabla^2 \phi_j({\bf r},t) + w({\bf r},t) \phi_j({\bf r},t)   \label{diff4}
\end{equation}
where the field $w({\bf r},t)$ must be a real function from equation (\ref{w3}) since both $n({\bf r},t)$ and $U[n]$ are real. Equation (\ref{diff4}) is the time-dependent KS equation \cite{Gross2006}. Due to the unitarity of equation (\ref{diff4}), orbitals $\phi_j({\bf r})$ defined by equation (\ref{phi1}) must remain orthogonal if their initial values are orthogonal. The initial condition is an unknown set of orbitals $\phi_j({\bf r},0)$, that is, the time evolution of $n({\bf r},t)$ depends on the initial state. This initial state dependence matches the initial state dependence of the Runge-Gross theorem in TDDFT \cite{Runge1984}, and, for fermions, one is forced to choose linearly independent (or most conveniently, orthogonal) initial conditions on the orbitals in order to enforce the Pauli exclusion principle. This condition naturally arises in the ring polymer thermal picture of quantum particles -- see section \ref{sec-Pauli} for an explanation of the requirement of linearly independent (orthogonal) orbitals in the context of classical SCFT. Since our orbitals give the density expression (\ref{dens6}) and the KS equation (\ref{diff4}), as well as being orthonormal, we can, without loss of generality, always transform them, for convenience, into eigenfunctions of the KS Hamiltonian (the right hand side operator of (\ref{diff4})). From now on, we will treat them as eigenfunctions. In practice, one normally starts with an initial state which is a solution to time-independent KS theory, that is, the equilibrium ring polymer SCFT equations. The initial state dependence goes away in such cases and the time evolution depends solely on the density $n({\bf r},t)$ \cite{Gross2006}. See section \ref{sec-initial} for further discussion of TDDFT initial state dependence.

Equations (\ref{dens6}) and (\ref{diff4}) are the expressions of KS-TDDFT \cite{Gross2006}, and TDDFT is known to match the dynamic predictions of QM through the Runge-Gross theorem \cite{Runge1984}.\footnote{See footnote \ref{note2}.} This means that postulating a mathematical correspondence between imaginary time (trajectories in the inverse temperature parameter) and real time (trajectories in spacetime) is sufficient to produce both static and dynamic QM.\footnote{Exchange and correlation effects in many-body systems can be included through exchange-correlation functionals in DFT.} In the time-dependent case, the choice of $\vert\psi_f(\tau)\rangle=\vert\psi_i(\tau)\rangle$ to obtain a real value might seem to require the knowledge of the time evolution beforehand, but the application of the unitary operator $\hat{U}(\tau,t)$ in (\ref{WVtdep1}) cancels the $\tau$-dependence for this specific choice of final boundary condition. Alternatively, one may use asymmetric boundary conditions and average the resulting weak values over a complete orthogonal set of final states. This is equivalent to removing the final boundary condition (retaining only the initial state), and yields the density for all times, with no explicit  $\tau$-dependence -- see appendix \ref{app-EfW}. This approach is conceptually different and rather general, as it does not presume a specific final boundary condition. In this ensemble approach, a single initial condition together with averaging over final boundary conditions yields the density throughout the time interval of interest. Although no measurements are involved here, similar alternatives arise in post-selected weak measurements (i.e., with conditioning on a chosen final state). If choosing identical pre- and post-selection states at the time of an instantaneous weak interaction, the measured weak value reduces to the expectation value in the initial state, and the probability of post-selection is close to one. If instead one measures the weak values for all possible post-selections in a specified measurement basis (a complete orthogonal set of final states), then the expectation value in the initial state is recovered from weighting these weak values by the corresponding post-selection probabilities. However, in an actual experiment, a weak interaction without post-selection already measures the expectation value in the initial state, showing that averaging over all post-selections is equivalent to bypassing post-selection. Independently of one's preferred approach to recover a real density, in the next section \ref{sec-finiteT}, we show that the model of classical ring polymers still robustly holds for time-dependent QM, but along the thermal axis rather than the real-time axis.

\subsection{Finite temperature dynamics}  \label{sec-finiteT}
To show that the ring polymer model of quantum particles is valid for time-dependent QM as well as for time-independent cases, it is necessary to show that the time and temperature-dependent density $n({\bf r},\beta,t)$ can be expressed in terms of a ring polymer propagator $\tilde{q}({\bf r},{\bf r},\beta,t)$.\footnote{Note that for ring polymers, propagators have two arguments of ${\bf r}$, giving the probability of starting at ${\bf r}$ and then returning to ${\bf r}$. This is on contrast to propagators for linear polymers that only have one ${\bf r}$ argument. See appendix \ref{app-polymers}.} In other words, we should be able to phrase QM in terms of closed imaginary time trajectories for time-dependent cases as well as time-independent situations. This was already done in reference \onlinecite{Thompson2022} by postulating the dynamics (\ref{diff4}) and using results of the Keldysh formalism \cite{vanLeeuwen2006b}. In section \ref{sec-TDDFT}, we showed how to obtain equation (\ref{diff4}) directly from the assumption of an imaginary time rather than postulating the dynamics. Now we will show that $n({\bf r},\beta,t)$ is proportional to $\tilde{q}({\bf r},{\bf r},\beta,t)$ without the Keldysh formalism or further assumptions. 

A temperature-and time-dependent density $n({\bf r},\beta,t)$ requires a temperature- and time-dependent KS potential, $w({\bf r},\beta,t)$, since the potential is found self-consistently from the density via equation (\ref{w3}). Since the time-dependent KS equation (\ref{diff3}) or (\ref{diff4}) depends on the potential, which is now a function of temperature $w({\bf r},\beta,t)$, the orbitals must also depend on temperature, $\phi_j({\bf r},\beta,t)$. The temperature-\emph{independent} orbitals are normalized, which means that our naive temperature-dependent orbitals cannot be if we continue with strict non-fractional occupancies. This is because at non-zero temperature, we don't just sum over the moduli of $N$ orbitals, but we must sum, in principle, to infinity, that is, over all possibly occupied orbitals.\footnote{Temperature-independent TDDFT also deals with non-ground state evolutions, but the initial state is almost exclusively chosen to be the $T=0$ case.} If the temperature-dependent orbitals were all normalized and simultaneously had non-fractional occupancies, the sum over the moduli of the orbitals would exceed the particle number $N$. Therefore, to reinstate normalization to our temperature-dependent orbitals, we separate out a fractional occupancy $f_j(\beta)$ for each orbital, giving a temperature-and time-dependent density expression of 
\begin{equation}
n({\bf r},\beta,t) = \sum_j f_j(\beta) |\phi_j({\bf r},\beta,t)|^2 .   \label{dens7}   
\end{equation}
For fermions in the canonical ensemble, the form of the occupancy $f_j(\beta)$ is known \cite{Giraud2018}. In the thermodynamic limit, it becomes the same as the Fermi-Dirac distribution where the chemical potential is fixed by the particle number $N$.\footnote{The \emph{canonical} ensemble expression for $f_j(\beta)$ is exact even for small numbers of particles and low temperatures, whereas the Fermi-Dirac distribution becomes inaccurate \cite{Barghathi2020}.} Note that, just like in the temperature-independent case, our orthonormal temperature-dependent orbitals can always be transformed without loss of generality into eigenfunctions of the \emph{temperature-dependent} KS Hamiltonian. We will label the corresponding eigenvalues at the time $t=0$ as $E_j^0$. We can now show that (\ref{dens7}) can be written in a form that is proportional to $\tilde{q}({\bf r},{\bf r},\beta,t)$.

Let us define the quantity 
\begin{equation}
\tilde{q}({\bf r}_0,{\bf r},\beta,t) \equiv \sum_{j=1}^N (-1)^{j-1} Z_{N-j}(\beta)\left[\sum_k e^{-j\beta E_k^0} \phi_k^*({\bf r}_0,\beta,t) \phi_k({\bf r},\beta,t) \right]  \label{q1}
\end{equation}
where $Z_N$ obeys the recursion relation 
\begin{equation}
Z_N(\beta) = \frac{1}{N} \sum_{j=1}^N (-1)^{j-1} \tilde{Q}_1(j\beta) Z_{N-j}(\beta)  \label{q2}
\end{equation}
with $Z_0=1$ by definition and
\begin{eqnarray}
\tilde{Q}_N(\beta) &\equiv& \int d{\bf r} \tilde{q}({\bf r},{\bf r},\beta,t) \nonumber \\
&=& \sum_{j=1}^N (-1)^{j-1} Z_{N-j}(\beta)\left(\sum_k e^{-j\beta E_k^0} \right)  . \label{tQ1}
\end{eqnarray}
We have used the orthonormality of the $\phi_k({\bf r},\beta,t)$ to simplify equation (\ref{tQ1}), and it will be shown presently that $\tilde{Q}_N(\beta) = N Z_N(\beta)$. The choice of the definition (\ref{q1}) and the form of recursion relation is based on the canonical ensemble formulas of Borrmann and Franke \cite{Borrmann1993} and Giraud et al. \cite{Giraud2018}. For a single particle $N=1$, equation (\ref{tQ1}) becomes just the single particle partition function
\begin{equation}
\tilde{Q}_1(\beta) = \sum_k e^{-\beta E_k^0}  \label{tQ2}
\end{equation}
so we can write (\ref{tQ1}) as 
\begin{equation}
\tilde{Q}_N(\beta) = \sum_j (-1)^{j-1} Z_{N-j}(\beta) \tilde{Q}_1(j\beta) .  \label{tQ3}
\end{equation}
Comparing (\ref{q2}) and (\ref{tQ3}) shows that $\tilde{Q}_N(\beta) = N Z_N(\beta)$ as expected. With these definitions we get
\begin{eqnarray}
N \frac{\tilde{q}({\bf r},{\bf r},\beta,t)}{\tilde{Q}_N(\beta)} &=&  \frac{N\sum_{j=1}^N (-1)^{j-1} Z_{N-j}(\beta)\left[\sum_k e^{-j\beta E_k^0} |\phi_k({\bf r},\beta,t) |^2\right]}{N Z_N(\beta)} \nonumber \\
&=& \sum_k\left[\frac{1}{Z_N(\beta)} \sum_{j=1}^N (-1)^{j-1} e^{-j\beta E_k^0} Z_{N-j}(\beta) \right] |\phi_k({\bf r},\beta,t) |^2  \nonumber \\
&=& \sum_k f_k(\beta) |\phi_k({\bf r},\beta,t) |^2 \label{tQ4}
\end{eqnarray}
where 
\begin{equation}
f_k(\beta) \equiv \frac{1}{Z_N(\beta)} \sum_{j=1}^N (-1)^{j-1} e^{-j\beta E_k^0} Z_{N-j}(\beta) .  \label{fCE}
\end{equation}
Equation (\ref{fCE}) is exactly the canonical ensemble occupation number formula given by Giraud et al. \cite{Giraud2018}, which becomes the Fermi-Dirac distribution in the thermodynamic limit. Equation (\ref{tQ4}) is identical to the time-dependent thermal DFT density result (\ref{dens7}), so by comparison we can say  
\begin{equation}
n({\bf r},\beta,t) = N \frac{\tilde{q}({\bf r},{\bf r},\beta,t)}{\tilde{Q}_N(\beta)} .  \label{dens8}
\end{equation}
This is to say, at each time $t$, it is possible to define a real-valued ring polymer propagator $\tilde{q}({\bf r},{\bf r},\beta,t)$ defined through (\ref{q1}) that is proportional to the density. This proves that a ring polymer model in imaginary time is compatible with time-dependent, \emph{finite temperature} QM and arises solely from the postulate of the mathematical correspondence between cyclic imaginary time and real time.

\section{Discussion}  \label{sec-Discussion}
A number of issues that arise in standard presentations of TDDFT also occur in this alternative derivation. These include causality, initial state dependence, and $v$-representability. In addition, we will review how the Pauli exclusion principle and fermion exchange can be included in a different way than in standard TDDFT. We now discuss each of these aspects.

\subsection{Causality}
In common with standard derivations of TDDFT, one can obtain the field $w({\bf r},t)$ through functional differentiation as in equation (\ref{w3}):
\begin{displaymath}
w({\bf r},t) = \frac{\delta U[n]}{\delta n({\bf r},t)} 
\end{displaymath}
where the factors $i$ in (\ref{w3}) have been dropped since the field must be real, as mentioned previously. From (\ref{w3}), we could also write
\begin{equation}
\frac{\delta w({\bf r},t)}{\delta n({\bf r}^\prime,t^\prime)} = \frac{\delta^2 U[n]}{\delta n({\bf r},t)\delta n({\bf r}^\prime,t^\prime)}  .  \label{kernel}
\end{equation}
To respect causality, the quantity on the left hand side of equation (\ref{kernel}) should only depend on densities $n({\bf r}^\prime,t^\prime)$ from \emph{earlier} times $t^\prime < t$. However the right hand side of (\ref{kernel}) is symmetric under interchange of $t$ and $t^\prime$, so it seems this equation must be inconsistent. To resolve this, note that we have already baked causality into the density $n({\bf r},t)$ by setting the time-evolved initial state as the final boundary condition at time $\tau$. With this choice, the information provided by the second boundary condition becomes redundant with that of the initial boundary condition since the time evolution is unitary. The density expressions (\ref{dens5}), (\ref{dens6}) and (\ref{dens8}) are therefore calculated using only the time-dependent KS equation (\ref{diff4}) for times $[0,t]$; information from the interval $(t,\tau]$ disappears (as $\langle \psi_i(\tau)\vert \hat{U}(\tau,t)=\langle \psi_i(t)\vert$), so the $(t,\tau]$ interval does not need to be incorporated and equation (\ref{kernel}) does not violate causality. For arbitrary final boundary conditions, the description is not causal; we have conditional evolution in this case corresponding to pre- and post-selection for the quantum weak value formula (\ref{weak2}) as presented in section \ref{sec-weak}. This resolution is essentially the logic of Vignale in disguise \cite{Vignale2008}.

\subsection{Initial state dependence}  \label{sec-initial}
The Runge-Gross theorem guarantees a one-to-one mapping between many-particle wave functions and single particle densities $n({\bf r},t)$ in TDDFT, \emph{for a given many-body initial state} $\psi_0({\bf r}_1, \cdots,  {\bf r}_N)$. For time-independent DFT, the Hohenberg-Kohn and Mermin theorems allow one to work just with the single particle density, but in TDDFT one needs, in general, both the time-dependent density \emph{and} the wave function at the initial time $t=0$ \cite{Gross2006}. In practice, if the system at the initial time is a solution to time-independent KS-DFT, then there is no initial state dependence and one can work solely in terms of $n({\bf r},t)$ for the dynamics \cite{Gross2006}. 

This can be understood in terms of ring polymer SCFT. For polymers, the initial density $n({\bf r},\beta)$ is all that is needed to solve the SCFT equations (\ref{dens1})--(\ref{diff1}) for an equilibrium system. For a non-equilibrium system however, there may be many different unrelaxed polymer configurations that could give the same density, and the equations (\ref{dens1})--(\ref{diff1}) no longer hold. One would need to know the specific microstate of the system for a given initial density, that is, all the out-of-equilibrium polymer configurations for that density would need to be known to specify the initial conditions. Likewise, in mapping from (\ref{dens1})--(\ref{diff1}) to the dynamic equations (\ref{dens3})--(\ref{diff3}), a full specification of the initial conditions $\psi_0({\bf r}_1, \cdots  {\bf r}_N)$ would be needed for densities $n({\bf r},0)$ that are not solutions to the equilibrium equations at initial times to unambiguously calculate the dynamics for $n({\bf r},t)$. 

\subsection{$v$-representability}
For a system with a given external potential, the theorems of DFT guarantee that there is a unique density that corresponds to that potential. It is this that allows one to work in terms of the one-body density $n({\bf r},t)$ (for the time-dependent case) rather than the many-body wave function. However, for an arbitrary density $n({\bf r},t)$, there is no guarantee that there exists any external potential that generates that density. This is the $v$-representability problem. In KS-DFT, including TDDFT, one assumes there exists a fictional non-interacting system subject to an artificial external potential which generates the exact same density $n({\bf r},t)$ as the real interacting system subject to the actual external potential. Again, there is no guarantee that such an artificial external potential exists, and this is the non-interacting $v$-representability problem. 

In the SCFT derivation for equilibrium polymers, no $v$-representability problem is discussed since the SCFT equations are derived from first principles in a finite temperature ensemble and mean-field framework without the use of the theorems of DFT. For time-independent quantum systems then, the KS-DFT equations are derived through SCFT in a way that is, at first glance, apparently free from the $v$-representability issue. This however assumes an exchange-correlation functional can be found that incorporates effects beyond the mean field. Likewise, mapping from cyclic imaginary time to non-cyclic real time should preserve this feature such that TDDFT derived this way should also be, in the same limited ensemble and mean-field sense, free of $v$-representability issues.\footnote{Levy-Lieb constrained search approaches to DFT likewise avoid aspects of the  $v$-representability problem, but not the overall problem -- see footnote \ref{note1}.} 

Another exception to the generality of this statement comes from the fact that the first principles SCFT derivation uses a functional Taylor series expansion \cite{Das1993}, and so one is assuming that the functional derivatives exist. This is analogous to the original derivation of the Runge-Gross theorem which uses a Taylor series expansion in time and so requires a potential that is analytic at the initial time \cite{Runge1984, Olevano2018}. In any case, the extension to finite temperatures, given in section \ref{sec-finiteT} and the inherent ensemble nature of polymer SCFT, reduces possible issues with $v$-representability. Deeper considerations of $v$-representability are beyond the scope of this article, and the reader is referred to references \onlinecite{vanLeeuwen1999, Ruggenthaler2015}.

\subsection{The Pauli exclusion principle} \label{sec-Pauli}
In practical calculations, the effects of the Pauli exclusion principle can be included as a Pauli contribution through the exchange-correlation term in DFT. This does not however provide any model for Pauli effects consistent with the classical statistical mechanical derivation of SCFT based on ring polymers. If this picture of quantum particles is followed rigorously -- that quantum particles are thermal ring polymers, that is, they are cyclic trajectories in imaginary time -- then we must impart all the properties of time on imaginary time (except of course, cyclicity and complexity). In this we follow Feynman's original development \cite{Feynman1953b}: if trajectories in cyclic imaginary time indeed behaves like those in real time, then different particles cannot exist at the same position and imaginary time just as they cannot exist at the same position and time. Feynman did not allow such overlaps, and this leads to shell structure in atoms and molecules that is known to be a consequence of the Pauli exclusion principle \cite{Thompson2020, Kealey2024, Sillaste2022, LeMaitre2023a}. More than this, in reference \onlinecite{Kealey2024}, exact Fermi-Dirac statistics were incorporated into the partition functions of ring polymers representing quantum particles in order to model electrons in a calculation of the beryllium atom. It was shown that these partition functions are exactly equivalent to classical partition functions for imaginary time trajectories that obey excluded volume, including topological effects associated with rings. This first principles calculation of the shell structure of the beryllium atom with no free parameters showed excellent agreement with Hartree-Fock theory, with the only approximation in addition to ignoring classical correlations coming from the neglect of electron self-interactions. This is despite not using an orbital approach, where orthogonal orbitals would automatically give an antisymmetric wave function if the many-particle wave function is formed as a Slater determinant. In fact, the statistics of ring polymers with no overlaps, as explained in reference \onlinecite{Kealey2024}, are exactly those given in section \ref{sec-finiteT}, with equation (\ref{q2}) being exactly expressible  as a Slater determinant \cite{Zhao2020}. So any ring polymers representing fermions have to obey these no-overlap statistics, effectively enforcing orthogonal initial conditions on the orbitals. More evidence in support of this classical imaginary time excluded volume interpretation of the Pauli exclusion principle, including scaling theory arguments, further SCFT calculations, and constraints on the electron density, are reviewed in reference \onlinecite{Kealey2024}. Thus the mapping between imaginary time and real time provides not just the dynamics of TDDFT as shown in the previous sections, but it also automatically includes the effects of fermion exchange and the Pauli exclusion principle. 

\section{Summary} \label{sec-Conclusions}
The partition function of a ring polymer type particle in thermal-space (imaginary time) is given by equation (\ref{part1}). The SCFT equations that are equivalent to time-independent KS-DFT arise from this partition function, and so all predictions of time-independent QM are obtained through the Hohenberg-Kohn and Mermin DFT theorems. If cyclic imaginary time is defined to mathematically correspond exactly to linear real time, except for the cyclicity and complexity, then the SCFT equations automatically give dynamic equations which we have shown to be equivalent to those of TDDFT. Through the Runge-Gross theorem, the assumption of the isomorphism between equations involving time and those involving cyclic imaginary time produces the same predictions as time-dependent QM.\footnote{See footnote \ref{note2}.} Our development respects causality, and like other TDDFT derivations, an initial state dependence remains, as do some aspects of the $v$-representability problem. We have shown that the dynamic results may be generalized to give us an expression for the temperature- and time-dependent single particle density $n({\bf r},\beta,t)$ as equation (\ref{dens7}), and, in section \ref{sec-finiteT} we have shown that this density can be expressed as proportional to a ring polymer propagator, given by equation (\ref{dens8}). More generally, the formula for the quantum weak value also arises spontaneously from the assumption of this mathematical correspondence. 

This real time $\leftrightarrow$ cyclic imaginary time correspondence gives rise to features that have been discussed in other papers, such as boson exchange \cite{Thompson2023}, entanglement \cite{Thompson2025}, the measurement problem, quantum kinetic energy, the uncertainty principle, the stability of atoms \cite{Thompson2019, Thompson2022}, molecular bonding \cite{Sillaste2022}, tunnelling, the double slit experiment and geometric phase including the Aharonov-Bohm effect \cite{Thompson2023}, classical DFT in the classical limit \cite{Thompson2019} and Kaluza theory with classical electromagnetism \cite{Thompson2022, Kaluza1921, Wesson1997, Dinov2022}. For many of these features, cyclic imaginary time may be viewed merely as a mathematical convenient fiction, but for some of them, it is tempting to take a strong view of the duality between real time and imaginary time, that is, to contemplate whether imaginary time is an element of reality. 

While the reader may remain agnostic, it is possible to consider taking the ring polymer structure in thermal-space-time as an ontology for QM. If one had classical physics but no notion of quantum mechanics, then the single assumption of cyclic imaginary time goes a long way towards reproducing the dynamic predictions of QM, as well as the other features mentioned above. Regardless of the ontological nature of the definition of quantum particles as classical ring polymers in imaginary time, the current approach is able to explain many features of QM, and through the theorems of DFT, in principle, \emph{all} features of QM,\footnote{See footnote \ref{note2}.}  assuming of course an exact exchange-correlation functional could be found. As mentioned in section \ref{sec-Pauli}, both boson and fermion exchange and entanglement effects can be incorporated, in principle exactly, in the SCFT formalism without departing from the classical ontology (see references  \onlinecite{Thompson2023, Kealey2024, Thompson2025}) and the remaining correlations are all classical and so could be included using the same methods as in standard DFT. It would be surprising if such an approach had not been examined previously. In fact, in 1965, Feynman and Hibbs speculated that their path integral method might be considered as an alternative to working with wave functions \cite{Feynman1965}. Much earlier than this however, both David Bohm in 1952, and before him, Louis de Broglie in 1927, had suggested that particles in QM might have definite trajectories \cite{Goldstein2010}. By construction, the classical approach described here assumes particles exist and have objective trajectories, just as in classical statistical mechanics, and so it would be valuable, in future work, to explore to what extent the ring polymer method falls into the class of de Broglie-Bohm theories \cite{Deotto1998}.

\appendix
\section{SCFT for Linear and Ring Polymers} \label{app-polymers}
\subsection{SCFT for Polymers and Quantum Particles}
SCFT for polymers is based on a coarse-grained picture of macromolecules, where the atomic details are ignored since, due to the large size of the polymer molecules, these details are relatively unimportant to the physical behaviour. In a melt, that is, in an incompressible liquid state composed entirely of a single type of polymer, these large molecules are known to obey the statistics of a completely random walk. Although one might suppose that a self-avoiding random walk would be necessary, since the molecules obviously can't pass through themselves or each other, in fact, it is the non-self-avoiding walk that gives the correct statistics. To understand why this is so, see the explanations of de Gennes \cite{deGennes1979} or Matsen \cite{Matsen2006}, for example. Due to this completely random walk behaviour in the melt, the random walk statistics often are taken to be the default, baseline, description, with self-avoiding walks and polymer chains of various stiffnesses considered more specialized models. Polymers that are modelled as these completely flexible chains are called \emph{Gaussian threads}, and they are pictured as perfect space-curves with no internal structure beyond their extended nature. At all lengths scales, they appear as Gaussian threads, and so are fractals. This model must obviously break down for real polymers at small enough length scales, just as it would for quantum particles with internal structure. For example, Feynman originally applied the quantum-classical isomorphism between polymers and quantum particles to the helium atom, ignoring the internal structure of the atoms \cite{Feynman1953b}. For particles without any known internal structure however, such as electrons, the model may be considered formally exact. 

SCFT is also a mean field theory, but the large size of polymers suppresses fluctuations and so the mean field approximation is excellent for macromolecules. For quantum particles however, it is necessary to account for both correlations and exchange effects, for which fortunately there is, through DFT, a large body of knowledge. The mapping between SCFT and DFT means that the same exchange-correlations functionals used in one can be used in the other. On the other hand, if one wants to preserve a classical correspondence for all terms, then exchange and entanglement effects can be included, in principle exactly, using ideas discussed elsewhere \cite{Thompson2023, Thompson2025, Kealey2024}, leaving only classical correlations, which can be handled in the regular DFT way.

\subsection{The SCFT Equations}
The main governing equation for SCFT is the diffusion equation\footnote{We have left the diffusion constant as $\hbar^2/2m$ in equation (\ref{A_diff1}) for internal consistency within this paper, but for actual polymer studies, the diffusion constant is related to the radius of gyration of the polymer.}
\begin{equation}
\frac{\partial q({\bf r}_0,{\bf r},s)}{\partial s} = \frac{\hbar^2}{2m} \nabla^2 q({\bf r}_0,{\bf r},s) - w({\bf r},s) q({\bf r}_0,{\bf r},s)    \label{A_diff1}
\end{equation}
where $s$ is a variable that parametrizes the contour of the polymer running from $s=0$ to $s=\beta$. While this equation can be derived rigorously from the path integral (\ref{part1}) -- see appendix B of reference \onlinecite{Helfand1975} -- it is easier to understand it heuristically. Since polymers are known to follow random walks in the melt, it is not surprising that a Gaussian thread polymer will conform to the diffusion equation, the solution of which is a random walk. Therefore, if one end of a linear polymer $s=0$ is known to be at a position ${\bf r}_0$, then $q({\bf r}_0,{\bf r},s^\prime)$ gives the unnormalized probability of the random walk arriving at position ${\bf r}$ after propagating from the $s=0$ end to the contour value $s=s^\prime$. The initial condition, corresponding to the statement that, ``one end of a linear polymer $s=0$ is known to be at a position ${\bf r}_0$'', is simply
\begin{equation}
q({\bf r}_0,{\bf r},0) = \delta({\bf r}-{\bf r}_0)  . \label{A_init1}
\end{equation}
The second term on the right hand side of equation (\ref{A_diff1}) involving $w({\bf r},s)$ modifies the diffusion equation to favour or suppress random walks into spatial regions where internal or external interactions are present. For example, an external field could confine the polymer, or interactions with other polymers could cause attraction or repulsion. 

For linear polymers where the two ends of the polymer are not symmetric, a separate propagator $q^\dag ({\bf r}^\prime_0,{\bf r},s)$ can be defined that runs from the $s=\beta$ end to $s=0$.\footnote{See footnote \ref{foot-dag}.} It obeys the same diffusion equation (\ref{A_diff1}) except with the left hand side multiplied by $-1$ to capture the reverse propagation direction. It obeys the initial condition 
\begin{equation}
q^\dag({\bf r}_0^\prime,{\bf r},\beta) = \delta({\bf r}-{\bf r}_0^\prime)  \label{A_init1b}
\end{equation}
where ${\bf r}_0^\prime$ is the known starting position of the $s=\beta$ end. The total unnormalized probability of finding a piece of the polymer $s$ at location ${\bf r}$ would therefore be the product of the likelihoods that both the $s=0$ end and the $s=\beta$ end propagate to the same point ${\bf r}$ for the contour value $s$. This would be the unnormalized conditional ``density'' for the segment $s$:
\begin{equation}
n({\bf r}_0,{\bf r}^\prime_0,{\bf r},s) \propto q({\bf r}_0,{\bf r},s) q^\dag({\bf r}^\prime_0,{\bf r},s)  . \label{A_dens1} 
\end{equation}
The normalization factor would include the integral over both initial positions ${\bf r}_0$ and ${\bf r}^\prime_0$, and the common final position ${\bf r}$:
\begin{equation}
Q(\beta) = \frac{1}{V} \iiint q({\bf r}_0,{\bf r},s) q^\dag({\bf r}^\prime_0,{\bf r},s) d{\bf r}_0 d{\bf r}^\prime_0 d{\bf r}      \label{A_Q1}  
\end{equation}
where we have included a factor $1/V$ to make $Q(\beta)$ dimensionless since the fundamental propagators $q({\bf r}_0,{\bf r},s)$ and $q^\dag({\bf r}_0^\prime,{\bf r},s)$ both have dimensions of inverse volume. The self-consistency of the equations is completed by relating the field $w({\bf r},s)$ to the density through equation (\ref{w1}) of section \ref{sec-Theory}. This is determined by the external and internal potentials of the system being modelled, and is not relevant to the rest of this appendix.

\subsection{Linear Polymers}
For linear polymers, it is possible to integrate out the ${\bf r}_0$ and ${\bf r}_0^\prime$ dependences. Defining the dimensionless propagator
\begin{equation}
\bar{q}({\bf r},s) \equiv \int q({\bf r}_0,{\bf r},s) d{\bf r}_0  \label{qbar}
\end{equation}
simplifies the modified diffusion equation to 
\begin{equation}
\frac{\partial \bar{q}({\bf r},s)}{\partial s} = \frac{\hbar^2}{2m} \nabla^2 \bar{q}({\bf r},s) - w({\bf r},s) \bar{q}({\bf r},s)   \label{A_diff2}
\end{equation}
which, dropping bars, is equation (\ref{diff1}) of section \ref{sec-Theory}. A similar propagator $\bar{q}^\dag({\bf r},s)$ can be defined by integrating over the ${\bf r}_0^\prime$ initial positions of the $s=\beta$ end of the polymer, and would obey equation (\ref{A_diff2}) with the left hand side multiplied by $-1$. The initial conditions (\ref{A_init1}) and (\ref{A_init1b}) become
\begin{eqnarray}
\bar{q}({\bf r},0) &=& 1  \label{A_init2}  \\
\bar{q}^\dag({\bf r},\beta) &=& 1  \label{A_init2b}
\end{eqnarray}
which are equations (\ref{init1}) and (\ref{init1b}) of section \ref{sec-Theory}. Equation (\ref{A_Q1}) becomes
\begin{equation}
\bar{Q}(\beta) = \frac{1}{V} \int \bar{q}({\bf r},s) \bar{q}^\dag({\bf r},s) d{\bf r}      \label{A_Q2}  
\end{equation}
matching equation (\ref{Q1}). The density will be the conditional expression (\ref{A_dens1}) integrated over ${\bf r}_0$ and ${\bf r}^\prime_0$ and normalized by (\ref{A_Q2})
\begin{equation}
\bar{n}({\bf r},s) =  \frac{n_0}{\bar{Q}(\beta)} \bar{q}({\bf r},s) \bar{q}^\dag({\bf r},s) \label{A_dens2} 
\end{equation}
where we have included a factor of the overall uniform density $n_0 = N/V$ to multiply this single particle density by the $N$ mean field particles in the system, giving the density expression (\ref{dens1}) of section \ref{sec-Theory}. 

\subsection{Ring Polymers}
While the mapping between SCFT and quantum dynamics uses linear polymers due to the open trajectories of quantum particles in the time domain, the mapping between SCFT and equilibrium quantum particles must use closed trajectories, that is, ring polymers, to reflect the closed loops in the imaginary time domain. For ring polymers, we are interested in the propagators $q({\bf r}_0,{\bf r},s)$ and $q^\dag({\bf r}_0,{\bf r},s)$ that each start from the \emph{same} initial position ${\bf r}_0$ at an arbitrary point on the ring assigned the value $s=0$ and which also corresponds to $s=\beta$, forming a closed loop. For this situation, it is no longer possible to integrate over the initial position as in equation (\ref{qbar}), and so the full diffusion equation (\ref{A_diff1}) must be solved for all possible ${\bf r}_0$ values. Equation (\ref{A_Q1}) for $Q(\beta)$ can be written as
\begin{equation}
Q(\beta) = \iint q({\bf r}_0,{\bf r},s) q^\dag({\bf r}_0,{\bf r},s) d{\bf r} d{\bf r}_0    \label{A_Q3}  
\end{equation}
where diffusion over $(0,s)$ starting from ${\bf r}_0$ meets up at ${\bf r}$ with diffusion in the other direction over $(\beta,s)$ also starting from ${\bf r}_0$ to form a ring. The cyclic condition removes one free volume integration, so $Q(\beta)$ is dimensionless without the explicit $1/V$ normalization factor present in (\ref{A_Q1}). Since $s$ merely labels an arbitrary point on a closed contour, $Q(\beta)$ is independent of $s$ despite the appearance of $s$ on the right hand side of (\ref{A_Q3}). Therefore, we can choose the value of $s=\beta$ and use the initial condition (\ref{A_init1b}) to simplify (\ref{A_Q3}) to
\begin{equation}
Q(\beta) = \int q({\bf r},{\bf r},\beta) d{\bf r} .   \label{A_Q4}  
\end{equation}
One could equally well phrase (\ref{A_Q4}) in terms of $q^\dag({\bf r},{\bf r},0)$ using $s=0$ and (\ref{A_init1}). The conditional density expression (\ref{A_dens1}) becomes 
\begin{equation}
n({\bf r},s) =  \frac{N}{Q(\beta)} \int q({\bf r}_0,{\bf r},s) q^\dag({\bf r}_0,{\bf r},s) d{\bf r}_0 \label{A_dens3} 
\end{equation}
after normalization. Again, if we are interested only in the value of the density at a given inverse temperature $s=\beta$, then using the initial condition (\ref{A_init1}), we can simplify equation (\ref{A_dens3}) to
\begin{equation}
n({\bf r},\beta) =  \frac{N}{Q(\beta)} q({\bf r},{\bf r},\beta)  .  \label{A_dens4} 
\end{equation}

\section{Recovering the Unconditioned Expectation Value by Averaging over Final Boundary Conditions} \label{app-EfW}

In section \ref{sec-TDDFT}, we recovered the real TDDFT density by specifically choosing the time-evolved initial state for the final boundary condition imposed at time $\tau$. As a result, the final boundary condition became effectively redundant, making it possible to compute the density as an initial-value problem, where weak values are no longer involved in practice. In this appendix, we show conceptually how to recover the same unconditioned expectation value that depends only on the initial state, in an alternative, rather general, construction that keeps arbitrary final boundary states at time $\tau$ but averages the related weak values over a complete set of orthogonal final states. We will follow the explanation given by Dressel and Jordan \cite{Dressel2012}. 

From Section \ref{sec-weak}, we had defined the unnormalized functions $\psi_i({\bf r},t) \equiv q({\bf r},it)/\sqrt{V}$ and $\psi_f^*({\bf r},t) \equiv q^\dagger({\bf r},it)/\sqrt{V}$ which allowed us to write the expression for the weak value in Dirac notation as equation (\ref{weak2}): 
\begin{equation}
\mathcal{O}_w^{(f)} (t) = \frac{\braket{\psi_f(t)|\hat{\mathcal{O}}|\psi_i(t)}}{\braket{\psi_f(t)|\psi_i(t)}} 
  \label{Eweak3}
\end{equation}
where (\ref{WVtdep1}) shows that $\vert\psi_i(t)\rangle=\hat{U}(t,0)\vert\psi_i(0)\rangle$ is the state propagated forward over a duration $t$ from the initial boundary condition $\vert\psi_i(0)\rangle$, while $\langle\psi_f(t)\vert=\langle\psi_f(\tau)\vert\hat{U}(\tau,t)$ is the state propagated backward over duration $\tau-t$ from the final boundary condition $\langle\psi_f(\tau)\vert$. As discussed at the end of section \ref{sec-weak}, the weak value (\ref{Eweak3}) depends implicitly on the time interval $\tau$ through the unitary operators involving $\tau$, as the final boundary condition $\langle\psi_f(\tau)\vert$ imposed at time $\tau$ is arbitrary.

We can take an appropriately weighted sum over a complete set of orthogonal final boundary states $\ket{\psi_f(\tau)}$ at an arbitrary final time  $\tau$ (similarly to the process of post-selecting at time $\tau$) according to
\begin{equation}
\sum_f\mathcal{P}(\psi_f(t)) \mathcal{O}_w^{(f)} (t) = \sum_f 
\frac{\left| \braket{\psi_f(t)|\psi_i(t)}\right|^2}{\braket{\psi_f(t)|\psi_f(t)}\braket{\psi_i(t)|\psi_i(t)}}
\frac{\braket{\psi_f(t)|\hat{\mathcal{O}}|\psi_i(t)}}{\braket{\psi_f(t)|\psi_i(t)}}   \label{Eweak4}
\end{equation}
where
\begin{equation}
\mathcal{P}(\psi_f(t)) \equiv 
\frac{\left| \braket{\psi_f(t)|\psi_i(t)}\right|^2}{\braket{\psi_f(t)|\psi_f(t)}\braket{\psi_i(t)|\psi_i(t)}}   \label{Eprob}
\end{equation}
is the probability of successfully post-selecting at time $t$ the backward-propagated state $\vert\psi_f(t)\rangle$ associated with the final boundary condition, given the forward-propagated initial condition $\vert\psi_i(t)\rangle$. Equation (\ref{Eweak4}) simplifies to 
\begin{eqnarray}
\sum_f\mathcal{P}(\psi_f(t)) \mathcal{O}_w^{(f)} (t) &=& \sum_f 
\frac{\braket{\psi_i(t)|\psi_f(t)} \braket{\psi_f(t)|\psi_i(t)}}{{\braket{\psi_f(t)|\psi_f(t)}\braket{\psi_i(t)|\psi_i(t)}}}
\frac{\braket{\psi_f(t)|\hat{\mathcal{O}}|\psi_i(t)}}{\braket{\psi_f(t)|\psi_i(t)}}   \nonumber \\
&=& \sum_f 
\frac{\braket{\psi_i(t)|\psi_f(t)}}{{\braket{\psi_f(t)|\psi_f(t)}\braket{\psi_i(t)|\psi_i(t)}}}
 \braket{\psi_f(t)|\hat{\mathcal{O}}|\psi_i(t)}  .  \label{Eweak5}
\end{eqnarray}
The closure relation removes all $\tau$-dependency on the right hand side:
\begin{equation}
\sum_f \frac{\ket{\psi_f(t)} \bra{\psi_f(t)}}{{\braket{\psi_f(t)|\psi_f(t)}}}  = \hat{\mathcal{I}}
 \label{Ecomplete}
\end{equation}
where $\hat{\mathcal{I}}$ is the identity operator. Equation (\ref{Eweak5}) becomes
\begin{eqnarray}
\braket{\mathcal{O}} &\equiv& \sum_f\mathcal{P}(\psi_f(t)) \mathcal{O}_w^{(f)} (t)  \nonumber  \\
&& =
\frac{ \braket{\psi_i(t)|\hat{\mathcal{O}}|\psi_i(t)}}{{\braket{\psi_i(t)|\psi_i(t)}}}
    \label{Eweak6}
\end{eqnarray}
which is the standard expectation value associated with the initial state evolved to time $t$, independent of $\tau$ and of any final boundary condition. In particular, it is independent of the basis that defined the set of final states. In brief, we have summed over a complete set of final boundary conditions at an arbitrary final time $\tau$, to get time evolution that depends solely on $t$ and a single initial condition. 

If the set of orthogonal final states at time $\tau$ includes the initial state at time $\tau$ (i.e. if $\vert\psi_f(\tau)\rangle=\vert\psi_i(\tau)\rangle$ for one $f$), the states propagated from the initial and final boundary conditions are always identical at time $t$. In that case, due to unitarity, a single probability (\ref{Eprob}) from the orthogonal set is non zero (the case $f=i$), so that a single weak value contributes to (\ref{Eweak4}), which reduces immediately to (\ref{Eweak6}). This relates the choice of $\vert\psi_f(\tau)\rangle=\vert\psi_i(\tau)\rangle$ performed in section \ref{sec-TDDFT} to the approach described here to get the real expectation value out of the weak value formulas (\ref{weak2}) and (\ref{WVtdep1}).

\begin{acknowledgments}
Y. C. is a Research Associate of the Fund for Scientific Research F.R.S.-FNRS. On behalf of all authors, the corresponding author states that there is no conflict of interest. There is no associated data.

\end{acknowledgments}

\bibliography{DFTbibliography}

@Article{Zhao2020,
  author = {{Y}u-{L}in Zhao and {C}hi-{C}hun Zhou and {W}en-{D}u Li and {W}u-{S}heng Dai},
  title = {Bose-like few-fermion systems},
    journal = {Physics Letters A},
  year    = {2020},
  volume  = {384},
  pages   = {126791},
}

@Article{Borrmann1993,
  author = {Peter Borrmann and Gert Franke},
  title = {Recursion formulas for quantum statistical partition functions},
    journal = {Journal of Chemical Physics},
  year    = {1993},
  volume  = {98},
  pages   = {2484-2485},
}

@Article{Giraud2018,
  author = {Olivier Giraud and Aur\'{e}lien Grabsch and Christophe Texier},
  title = {Correlations of occupation numbers in the canonical ensemble and application to a {B}ose-{E}instein condensate in a one-dimensional harmonic trap},
    journal = {Physical Review A},
  year    = {2018},
  volume  = {97},
  pages   = {053615},
}

@Article{Barghathi2020,
  author = {Hatem Barghathi and Jiangyong Yu and Adrian {D}el {M}aestro},
  title = {Theory of noninteracting fermions and bosons in the canonical ensemble},
    journal = {Physical Review Research},
  year    = {2020},
  volume  = {2},
  pages   = {043206},
}

@Article{Dressel2014b,
  author = {Justin Dressel and Mehul Malik and Filippo M. Miatto and Andrew N. Jordan and Robert W. Boyd},
  title = {Understanding quantum weak values:
{B}asics and applications},
    journal = {Reviews of Modern Physics},
  year    = {2014},
  volume  = {86},
  pages   = {307-316},
}

@Article{Dressel2014a,
  author = {Justin Dressel and Konstantin Y. Bliokh and Franco Nori},
  title = {Classical Field Approach to Quantum Weak Measurements},
    journal = {Physical Review Letters},
  year    = {2014},
  volume  = {112},
  pages   = {110407},
}

@Article{Dressel2015,
  author = {Justin Dressel},
  title = {Weak values as interference phenomena},
    journal = {Physical Review A},
  year    = {2015},
  volume  = {91},
  pages   = {032116},
}

@Article{Dressel2012,
  author = {Justin Dressel and Andrew N. Jordan},
  title = {Significance of the imaginary part of the weak value},
    journal = {Physical Review Letters},
  year    = {2012},
  volume  = {85},
  pages   = {012107},
}

@Article{LeMaitre2023a,
  author = {Phil A. LeMaitre and Russell B. Thompson},
  title = {Gaussian Basis Functions for an Orbital-Free-Related Density Functional Theory of Atoms},
    journal = {International Journal of Quantum Chemistry},
  year    = {2023},
  volume  = {123},
  pages   = {e27111},
}

@Article{LeMaitre2023b,
  author = {Phil A. LeMaitre and Russell B. Thompson},
  title = {On the Origins of Spontaneous Spherical Symmetry-Breaking in Open-Shell Atoms Through Polymer Self-Consistent Field Theory},
  journal = {Journal of Chemical Physics},
  year    = {2023},
  volume  = {158},
  pages   = {064301},
}

@Article{Kealey2024,
  author = {Malcolm A. Kealey and Phil A. LeMaitre and Russell B. Thompson},
  title = {Fermion exchange in ring polymer quantum theory},
  journal = {Physical Review A},
  year    = {2024},
  volume  = {109},
  pages   = {052819},
}

@Article{Thompson2025,
  author = {Russell B. Thompson},
  title = {Visualizing quantum entanglement in {B}ose-{E}instein condensates
without state vectors},
  journal = {International Journal of Theoretical Physics},
  year    = {2025},
  volume  = {64},
  pages   = {13},
}

@Article{Kaluza1921,
  author  = {Theodor Kaluza},
  title   = {Zum Unit\"{a}tsproblem der Physik},
  journal = {Sitz. Preuss. Akad. Wiss. Phys. Math. Kl.},
  year    = {1921},
  pages   = {966-972},
}

@Article{Wesson1997,
  author  = {James M. Overduin and Paul S. Wesson},
  title   = {{K}aluza-{K}lein Gravity},
  journal = {Phys. Rep.},
  year    = {1997},
  volume  = {283},
  pages   = {303-378},
}

@Book{Dinov2022,
  title     = {Data Science},
  publisher = {de Gruyter GmbH},
  year      = {2022},
  author    = {Ivo D. Dinov and Milen Velchev Velev},
  address   = {Berlin, Germany},
}

@Article{Ceperley1995,
  author  = {D. M. Ceperley},
  title   = {Path integrals in the theory of condensed helium},
  journal = {Reviews of Modern Physics},
  year    = {1995},
  volume  = {67},
  pages   = {279-355},
}

@Article{Zeng2014,
  author  = {Tao Zeng and Pierre-Nicholas Roy},
  title   = {Microscopic molecular superfluid response: theory and simulations},
  journal = {Reports on Progress in Physics},
  year    = {2014},
  volume  = {77},
  pages   = {046601},
}

@Article{Roy1999b,
  author  = {Pierre-Nicholas Roy and G. A. Voth},
  title   = {Feynman path centroid dynamics for {Fermi-Dirac} statistics},
  journal = {J. Chem. Phys.},
  year    = {1999},
  volume  = {111},
  pages   = {5303-5305},
}

@Article{Feynman1953b,
  author  = {Richard P. Feynman},
  title   = {Atomic Theory of the $\lambda$-Transition in Helium},
  journal = {Physical Review},
  year    = {1953},
  volume  = {91},
  pages   = {1291-1301},
}

@Book{Feynman1965,
  title     = {Quantum Mechanics and Path Integrals},
  publisher = {Dover Publications},
  year      = {1965},
  author    = {Richard P. Feynman and Albert R. Hibbs},
  address   = {Mineola NY},
}

@Article{Deotto1998,
  author  = {E. Deotto and G. C. Ghirardi},
  title   = {{B}ohmian Mechanics Revisited},
  journal = {Foundations of Physics},
  year    = {1998},
  volume  = {28},
  pages   = {1-30},
}

@Article{Althorpe2009,
  author  = {Jeremy O. Richardson and Stuart C. Althorpe},
  title   = {Ring-polymer molecular dynamics rate-theory in the deep-tunneling regime: Connection with semiclassical instanton theory},
  journal = {Journal of Chemical Physics},
  year    = {2009},
  volume  = {131},
  pages   = {214106},
}

@Article{Thompson2019,
  author  = {Russell B. Thompson},
  title   = {An alternative derivation of orbital-free density functional theory},
  journal = {Journal of Chemical Physics},
  year    = {2019},
  volume  = {150},
  pages   = {204109},
}

@Article{Thompson2020,
  author  = {Russell B. Thompson},
  title   = {Atomic shell structure from an orbital-free-related density-functional-theory {Pauli} potential},
  journal = {Physical Review A},
  year    = {2020},
  volume  = {102},
  pages   = {012813},
}

@Article{Hohenberg1964,
  author  = {P. Hohenberg and W. Kohn},
  title   = {Inhomogeneous Electron Gas},
  journal = {Physical Review},
  year    = {1964},
  volume  = {136},
  pages   = {B864-B871},
}

@Article{vonBarth2004,
  author  = {U. von Barth},
  title   = {Basic Density-Functional Theory --- an Overview},
  journal = {Physica Scripta},
  year    = {2004},
  volume  = {T109},
  pages   = {9-39},
}

@incollection{Goldstein2010,
  author  = {Sheldon Goldstein and Roderich Tumulka and Nino Zanghi},
  title   = {Bohmian Trajectories as the Foundation of Quantum Mechanics},
  editor = {Pratim Kumar Chattaraj},
  booktitle = {{Q}uantum {T}rajectories},
  publisher = {CRC Press, Taylor \& Francis},
  address = {New York},
  year    = {2011},
  pages   = {1-16},
  }

@Article{vanLeeuwen1999,
  author  = {Robert van Leeuwen},
  title   = {Mapping from Densities to Potentials in Time-Dependent Density-Functional Theory},
  journal = {Physical Review Letters},
  year    = {1999},
  volume  = {82},
  pages   = {3863-3866},
}

@incollection{vanLeeuwen2006b,
 author  = {Robert {van Leeuwen} and N. E. Dahlen and G. Stefanucci and C.-O. Almbladh and
 U. von Barth},
 title   = {Introduction to the {K}eldysh Formalism},
 editor = {Miguel A.L. Marques and Carsten A. Ullrich and Fernando Nogueira and Angel Rubio and Kieron Burke and Eberhard K. U. Gross},
  booktitle = {Time-Dependent Density Functional Theory},
  publisher = {Springer},
  address = {New York},
  year    = {2006},
  pages   = {33-57},
}

@incollection{Gross2006,
 author  = {E.K.U. Gross and K. Burke},
 title   = {[{DFT}] {B}asics},
 editor = {Miguel A.L. Marques and Carsten A. Ullrich and Fernando Nogueira and Angel Rubio and Kieron Burke and Eberhard K. U. Gross},
  booktitle = {Time-Dependent Density Functional Theory},
  publisher = {Springer},
  address = {New York},
  year    = {2006},
  pages   = {1-12},
}

@Article{Kim2012,
  author  = {Jaeup U. Kim and Yong - Biao Yang and Won Bo Lee},
  title   = {Self-Consistent Field Theory of {G}aussian Ring Polymers},
  journal = {Macromolecules},
  year    = {2012},
  volume  = {45},
  pages   = {3263-3269},
}

@Article{Qiu2011,
  author  = {Guojie Zhang and Zhongyong Fan and Yuliang Yang and Feng Qiu},
  title   = {Phase behaviors of cyclic diblock copolymers},
  journal = {Journal of Chemical Physics},
  year    = {2011},
  volume  = {135},
  pages   = {174902},
}

@Article{Vignale2001,
  author  = {Klaus Capelle and Giovanni Vignale},
  title   = {Nonuniqueness of the Potentials of Spin-Density-Functional Theory},
  journal = {Physical Review Letters},
  year    = {2001},
  volume  = {86},
  pages   = {5546-5549},
}

@Article{Vignale2002,
  author  = {Klaus Capelle and Giovanni Vignale},
  title   = {Nonuniqueness and derivative discontinuities in density-functional theories for current-carrying and superconducting systems},
  journal = {Physical Review B},
  year    = {2002},
  volume  = {65},
  pages   = {113106},
}

@Article{Vignale2008,
  author  = {Giovanni Vignale},
  title   = {Real-time resolution of the causality paradox of time-dependent density-functional theory},
  journal = {Physical Review A},
  year    = {2008},
  volume  = {77},
  pages   = {062511},
}

@Article{Vignale1987,
  author  = {Giovanni Vignale and Mark Rasolt},
  title   = {Density-Functional Theory in Strong Magnetic Fields},
  journal = {Physical Review Letters},
  year    = {1987},
  volume  = {59},
  pages   = {2360-2363},
}

@Article{DelleSite2014,
  author  = {Luigi {D}elle {S}ite},
  title   = {{L}evy–{L}ieb principle: The bridge between the electron density of Density Functional Theory and the wavefunction of Quantum {M}onte {C}arlo},
  journal = {Chemical Physics Letters},
  year    = {2014},
  volume  = {619},
  pages   = {148-151},
}

@Article{Levy1979,
  author  = {Mel Levy},
  title   = {Universal variational functionals of electron densities, first-order density matrices, and natural spin-orbitals and solution of the $v$-representability problem},
  journal = {Proceedings of the National Academy of Sciences},
  year    = {1979},
  volume  = {76},
  pages   = {6062-6065},
}

@Article{Lieb1983,
  author  = {Elliott H. Lieb},
  title   = {Density functionals for {C}oulomb systems},
  journal = {International Journal of Quantum Chemistry},
  year    = {1983},
  volume  = {24},
  pages   = {243-277},
}

@Article{Cohen2005,
  author  = {Morrel H. Cohen and Adam Wasserman},
  title   = {$N$-representability and stationarity in time-dependent density-functional theory},
  journal = {Physical Review A},
  year    = {2005},
  volume  = {71},
  pages   = {032515},
}

@Article{Ullrich2025,
  author  = {Carsten A. Ullrich},
  title   = {A snapshot of time-dependent density-functional theory},
  journal = {APL Computational Physics},
  year    = {2025},
  volume  = {1},
  pages   = {020901},
}

@Article{Aharonov1988,
  author  = {Yakir Aharonov and David Z. Albert and Lev Vaidman},
  title   = {How the Result of a Measurement of a Component of the Spin of a
Spin- 2 Particle Can Turn Out to be 100},
  journal = {Physical Review Letters},
  year    = {1988},
  volume  = {60},
  pages   = {1351-1354},
}

@Article{Aharonov1964,
  author  = {Yakir Aharonov and Peter G. Bergmann and Joel L. Lebowitz},
  title   = {Time Symmetry in the Quantum Process of Measurement},
  journal = {Physical Review},
  year    = {1964},
  volume  = {134},
  pages   = {B1410-B1416},
}

@Article{Steinberg2013,
  author  = {Aephraim Steinberg and Amir Feizpour and Lee Rozema and Dylan Mahler and Alex Hayat},
  title   = {In praise of weakness},
  journal = {Physics World},
  year    = {2013},
  volume  = {26},
  pages   = {35-40},
}

@Article{Svensson2013,
  author  = {Bengt E. Y. Svensson},
  title   = {Pedagogical Review of Quantum Measurement Theory with an Emphasis on Weak Measurements},
  journal = {Quanta},
  year    = {2013},
  volume  = {2},
  pages   = {18-49},
}

@Article{Lundeen2011,
  author  = {Jeff S. Lundeen and Brandon Sutherland and Aabid Patel and Corey Stewart and Charles Bamber},
  title   = {Direct measurement of the quantum wavefunction},
  journal = {Nature},
  year    = {2011},
  volume  = {474},
  pages   = {188-191},
}

@incollection{Matsen2006,
  author  = {Mark W. Matsen},
  title   = {Self-Consistent Field Theory and Its Applications},
  editor = {G. Gompper and M. Schick},
  booktitle = {Soft Matter, Volume 1: Polymer Melts and Mixtures},
  publisher = {Wiley-VCH},
  address = {Weinheim},
  year    = {2006},
  pages   = {87-178},
}

@incollection{Greenberger2009,
  author  = {Paul Busch and Pekka Lahti},
  title   = {Measurement Theory},
  editor = {Daniel Greenberger and Klaus Hentschel and Friedel Weinert},
  booktitle = {Compendium of Quantum Physics},
  publisher = {Springer-Verlag},
  address = {New York, USA},
  year    = {2009},
  pages   = {374-379},
}

@Article{Qiu2006,
  author  = {Yuliang Yang and Feng Qiu and Ping Tang and Hongdong Zhang},
  title   = {Applications of self-consistent field theory in polymer systems},
  journal = {Science in China: Series B Chemistry},
  year    = {2006},
  volume  = {49},
  pages   = {21-43},
}

@Article{Schmid1998,
  author  = {Friederike Schmid},
  title   = {Self-consistent-field theories for complex fluids},
  journal = {Journal of Physics: Condensed Matter},
  year    = {1998},
  volume  = {10},
  pages   = {8105-8138},
}

@Article{Mermin1965,
  author  = {N. David Mermin},
  title   = {Thermal Properties of the Inhomogeneous Electron Gas},
  journal = {Physical Review},
  year    = {1965},
  volume  = {137},
  pages   = {A1441-A1443},
}

@Article{Runge1984,
  author  = {Erich Runge and Eberhard K. U. Gross},
  title   = {Density-Functional Theory for Time-Dependent Systems},
  journal = {Phys. Rev. Lett.},
  year    = {1984},
  volume  = {52},
  pages   = {997-1000},
}

@Book{deGennes1979,
  title     = {Scaling Concepts in Polymer Physics},
  publisher = {Cornell University Press},
  year      = {1979},
  author    = {Pierre-Gilles {de Gennes}},
  address   = {Ithaca NY},
}

@Article{Edwards1965,
  author  = {S. F. Edwards},
  title   = {The statistical mechanics of polymers with excluded volume},
  journal = {Proceedings of the Physical Society},
  year    = {1965},
  volume  = {85},
  pages   = {613-624},
}

@Article{Matsen2002,
  author  = {Mark W. Matsen},
  title   = {The standard {G}aussian model for block copolymer melts},
  journal = {Journal of Physics: Condensed Matter},
  year    = {2002},
  volume  = {14},
  pages   = {R21-R47},
}

@Article{Habershon2013,
  author  = {Scott Habershon and David E. Manolopoulos and Thomas E. Markland and Thomas F. Miller},
  title   = {Ring-Polymer Molecular Dynamics: Quantum Effects in Chemical Dynamics from Classical Trajectories in an Extended Phase Space},
  journal = {Annual Review of Physical Chemistry},
  year    = {2013},
  volume  = {64},
  pages   = {387-413},
}

@Article{Videla2023,
  author  = {Pablo E. Videla and Victor S. Batista},
  title   = {An exact imaginary-time path-integral phase-space formulation of multi-time correlation functions},
  journal = {Journal of Chemical Physics},
  year    = {2023},
  volume  = {158},
  pages   = {094101},
}

@Book{Fredrickson2006,
  title     = {The Equilibrium Theory of Inhomogeneous Polymers},
  publisher = {Oxford University Press},
  year      = {2006},
  author    = {Glenn H. Fredrickson},
  address   = {New York, NY},
}

@Book{Fredrickson2023,
  title     = {Field-Theoretic Simulations in Soft Matter and Quantum Fluids},
  publisher = {Oxford University Press},
  year      = {2023},
  author    = {Glenn H. Fredrickson and Kris T. Delaney},
  address   = {Oxford, UK},
}

@Article{Dornheim2019,
  author  = {Tobias Dornheim},
  title   = {Fermion sign problem in path integral {M}onte {C}arlo simulations: Quantum dots, ultracold atoms, and warm dense matter},
  journal = {Physical Review E},
  year    = {2019},
  volume  = {100},
  pages   = {023307},
}

@Article{Sillaste2022,
  author  = {Spencer Sillaste and Russell B. Thompson},
  title   = {Molecular Bonding in an Orbital-Free-Related Density Functional Theory},
  journal = {Journal of Physical Chemistry A},
  year    = {2022},
  volume  = {126},
  pages   = {325-332},
}

@article{Thompson2022,
  author  = {Russell B. Thompson},
  title   = {An Interpretation of Quantum Foundations Based on Density Functional Theory and Polymer Self-Consistent Field Theory},
  journal = {Quantum Studies: Mathematics and Foundations},
year    = {2022},
  volume  = {9},
  pages   = {405-416},
}

@article{Thompson2023,
  author  = {Russell B. Thompson},
  title   = {A Holographic Principle for Non-Relativistic Quantum Mechanics},
  journal = {International Journal of Theoretical Physics},
year    = {2023},
  volume  = {62},
  pages   = {34:1-15},
}

@Article{Matsen1994b,
  author  = {Mark W. Matsen and Michael Schick},
  title   = {Stable and Unstable Phases of a Linear Multiblock Copolymer Melt},
  journal = {Macromolecules},
  year    = {1994},
  volume  = {27},
  pages   = {7157-7163},
}

@Book{Das1993,
  title     = {Field Theory: A Path Integral Approach},
  publisher = {World Scientific},
  year      = {1993},
  author    = {Ashok Das},
  address   = {River Edge NJ},
}

@Article{Helfand1975,
  author  = {Eugene Helfand},
  title   = {Theory of inhomogeneous polymers: {F}undamentals of the {G}aussian random-walk model},
  journal = {Journal of Chemical Physics},
  year    = {1975},
  volume  = {62},
  pages   = {999-1005},
}

@Article{Ruggenthaler2015,
  author  = {Michael Ruggenthaler and Markus Penz and Robert van Leeuwen},
  title   = {Existence, uniqueness, and construction of the density-potential mapping in time-dependent density-functional theory},
  journal = {Journal of Physics: Condensed Matter},
  year    = {2015},
  volume  = {27},
  pages   = {203202},
}

@incollection{Butterfield2021,
  author  = {Jeremy Butterfield},
  title   = {On Dualities and Equivalences between Physical Theories},
  editor = {Christian W\"{u}thrich and Baptiste Le Bihan and Nick Hugget},
  booktitle = {Philosophy Beyond Spacetime: Implications from Quantum Gravity},
  publisher = {Oxford {U}niversity {P}ress},
  address = {Oxford, UK},
  year    = {2021},
  pages   = {41-77},
}

@incollection{Olevano2018,
  author  = {Valerio Olevano},
  title   = {{TDDFT}, Excitations and Spectroscopy},
  editor = {Theo Woike and Dominik Schaniel},
  booktitle = {Structures on {D}ifferent {T}ime {S}cales},
  publisher = {{D}e {G}ruyter},
  address = {Berlin, Germany},
  year    = {2018},
  pages   = {22},
}

@article{Cormann2017,
	title = {Geometric description of modular and weak values in discrete quantum systems using the {Majorana} representation},
	volume = {50},
	issn = {1751-8113, 1751-8121},
	url = {https://iopscience.iop.org/article/10.1088/1751-8121/aa7639},
	doi = {10.1088/1751-8121/aa7639},
	number = {30},
	urldate = {2022-10-23},
	journal = {Journal of Physics A: Mathematical and Theoretical},
	author = {Cormann, Mirko and Caudano, Yves},
	month = jul,
	year = {2017},
	pages = {305302}
}

@article{Ferraz2022,
	title = {Geometrical interpretation of the argument of weak values of general observables in {N}-level quantum systems},
	volume = {7},
	issn = {2058-9565},
	url = {http://arxiv.org/abs/2202.11145},
	doi = {10.1088/2058-9565/ac8bf1},
	number = {4},
	urldate = {2022-10-23},
	journal = {Quantum Science and Technology},
	author = {Ferraz, Lorena Ballesteros and Lambert, Dominique L. and Caudano, Yves},
	month = oct,
	year = {2022},
	note = {arXiv:2202.11145 [quant-ph]},
	pages = {045028}
}

@article{Ferraz2025,
doi = {10.1088/1751-8121/add22a},
url = {https://doi.org/10.1088/1751-8121/add22a},
year = {2025},
month = {may},
publisher = {IOP Publishing},
volume = {58},
number = {20},
pages = {205303},
author = {Ballesteros Ferraz, Lorena and Lambert, Dominique and Caudano, Yves},
title = {Exploring weak value arguments and {B}argmann invariants in {N}-level quantum systems through the {M}ajorana symmetric representation},
journal = {Journal of Physics A: Mathematical and Theoretical}
}

@article{Nelson1966,
  title = {Derivation of the {S}chr\"odinger Equation from {N}ewtonian Mechanics},
  author = {Nelson, Edward},
  journal = {Phys. Rev.},
  volume = {150},
  issue = {4},
  pages = {1079--1085},
  numpages = {0},
  year = {1966},
  month = {Oct},
  publisher = {American Physical Society},
  doi = {10.1103/PhysRev.150.1079},
  url = {https://link.aps.org/doi/10.1103/PhysRev.150.1079}
}

@incollection{Bacciagaluppi2005,
  author    = {Bacciagaluppi, Guido},
  title     = {A Conceptual Introduction to {N}elson's Mechanics},
  booktitle = {Endophysics, Time, Quantum and the Subjective: Proceedings of the ZiF Interdisciplinary Research Workshop, 17--22 January 2005, Bielefeld, Germany},
  editor    = {Buccheri, Rosolino and Elitzur, Avshalom C. and Saniga, Metod},
  pages     = {367--388},
  publisher = {World Scientific},
  address   = {Singapore},
  year      = {2005},
  doi       = {10.1142/9789812701596_0020},
  isbn      = {9789812565099}
}

@article{BallesterosFerraz2024,
doi = {10.1088/2058-9565/ad420b},
url = {https://doi.org/10.1088/2058-9565/ad420b},
year = {2024},
month = {may},
publisher = {IOP Publishing},
volume = {9},
number = {3},
pages = {035029},
author = {Ballesteros Ferraz, Lorena and Martin, John and Caudano, Yves},
title = {On the relevance of weak measurements in dissipative quantum systems},
journal = {Quantum Science and Technology}
}

\end{document}